\documentclass[useAMS,usenatbib]{mn2e}
\usepackage{amssymb}
\usepackage{longtable}
\usepackage[pdftex]{graphicx}
\title[Optimal filtering to search for galaxy clusters]{Optimal filtering of optical and weak lensing data to search for galaxy clusters: application to the COSMOS field}
\author[Bellagamba et al.]{F. Bellagamba$^{1}$, M. Maturi$^{2}$, T. Hamana$^{3}$, M. Meneghetti$^{4,5}$, S. Miyazaki$^{3}$,
\newauthor
L. Moscardini$^{1,5}$\\
$^{1}$Dipartimento di Astronomia, Universit\`{a} di Bologna, Via Ranzani 1, 40127, Bologna, Italy\\
$^{2}$Zentrum f\"{u}r Astronomie, ITA, Universit\"{a}t Heidelberg, Albert-Ueberle-Str. 2, 69120, Heidelberg, Germany\\
$^{3}$National Astronomical Observatory of Japan, Mitaka, Tokyo 181-8588, Japan\\
$^{4}$INAF-Osservatorio Astronomico di Bologna, Via Ranzani 1, 40127, Bologna, Italy\\
$^{5}$INFN-National Institute for Nuclear Physics, Sezione di Bologna, Viale Berti Pichat 6/2, 40127, Bologna, Italy}

\begin{document}

\date{}

\pagerange{\pageref{firstpage}--\pageref{lastpage}} \pubyear{2009}

\maketitle

\label{firstpage}

\begin{abstract}

Galaxy clusters are usually detected in blind optical surveys via
suitable filtering methods. We present an optimal matched filter
which maximizes their signal-to-noise ratio by taking
advantage of the knowledge we have of their intrinsic physical
properties and of the data noise properties. In this paper we
restrict our application to galaxy magnitudes, positions
and photometric redshifts if available, and we also apply the filter
separately to weak lensing data. The method is suitable to be
naturally extended to a multi-band approach which could include not
only additional optical bands but also observables with different
nature such as X-rays. For each detection, the filter provides its
significance, an estimate for the richness and for the redshift
even if photo-z are not given. The provided analytical error
estimate is tested against numerical simulations. We finally apply
our method to the COSMOS field and compare the results with
previous cluster detections obtained with different methods. Our
catalogue contains 27 galaxy clusters with minimal threshold
at 3-sigma level including both optical and weak-lensing information.

\end{abstract}
%}

\section{Introduction}

Clusters of galaxies lie in the densest regions of the Universe. In the standard $\Lambda$CDM model, they form from the highest peaks of the primordial density fluctuations. Their abundance as a function of mass and redshift is an extremely important tool to determine cosmological parameters and the history of structure formation. It is therefore fundamental to build robust tools to detect galaxy clusters in different observational bands.

In optical surveys, clusters of galaxies can be succesfully identified via
overdensities of galaxies with respect to the mean field and via the
coherent distortion of background galaxies due to gravitational lensing. These two observables come together with the same data but have very different properties.
On one hand lensing analyses have the advantage of directly mapping dark
matter structures, but to be performed they need high-quality imaging data, and
their results are intrinsically very noisy. On the other hand, detecting clusters
through galaxy overdensities is much less demanding from an
observational point of view, because it does not require any shape measurement.
Moreover, if an optical cluster finding is performed on a lensing survey field, it
can be used as a useful tool to eliminate spurious detections.

A common approach in lensing is to convolve galaxy ellipticities with a suitable filter  to find the cluster signal beyond the noise induced by intrinsic
ellipticity of galaxies and cosmic shear by large-scale structures. Different filters have been proposed so far, including the Gaussian filter \citep{miy}, the aperture mass in the version
of \citet{b14}, its modifications by \citet{b13}, and the one proposed by \citet{hs}.
\citet{b7} proposed an optimal filter, which minimizes noises due to both the intrinsic ellipticity and the cosmic shear,
selecting the scales on which the cluster signal-to-noise ratio is expected to be maximum. A
comparison of the performance of the different filters used to find dark matter
halos in cosmological simulations has been presented in \citet{b9}. An application to GaBoDs data of the optimal filter is in \citet{b20}.

The idea of this work is to apply the linear filtering technique to
optical galaxy catalogues, possibly accounting for 3D
information, if available through photometric redshifts, to perform a combined analysis of galaxy overdensities and weak lensing data.

To build an optimal linear filter for galaxy overdensities, we
start from the pioneering approach proposed by \citet{b12}, who developed an optical
filter to analyse the data of the Palomar Distant Cluster Survey. Many authors
have proposed different modifications of the original Postman filter: for example
\citet{b5} introduced Poisson (instead of the Gaussian) statistics and a rough usage
of photometric redshifts; \citet{b1} allowed the radius of the filter to change
to adapt to the detected signal; \citet{b19} tuned the parameters using mock lightcones extracted from the Millennium Simulation \citep{spring}.

In this work, we construct a Postman-like filter as a linear filter, using the estimated richness of the cluster and
not the likelihood of the data as main object of analysis. This allows us
to keep the linearity with respect to the galaxy density and to calculate an error estimate analytically. We also introduce the information from
photometric redshifts in the procedure. We apply the filter to some
simulated catalogues to verify the consistency of the error estimates and to
assess the reliability of the cluster redshift determination obtained with our algorithm, even when galaxy
photometric redshifts are not available. The final step of this work is then the application of both optimal filtering algorithms (for optical and weak lensing data) to the COSMOS field \citep{cosmos}. Previous cluster detections on the same field were done by \citet{b2} combining optical and X-ray information, by \citet{b8} and \citet{b15} from optical data only, and by \citet{b4} and \citet{b21} via lensing analyses.

The paper is organized as follows. In Section 2, we show the idea, the
mathematical derivation and our implementation of the filter. We also
describe how to derive a redshift estimate for the detected galaxy
clusters, both with and without the information coming from photometric redshifts. In Section 3, we investigate the filter properties and test the algorithm using numerical simulations. In Section 4, we briefly summarize the properties of the optimal filter for weak lensing detections, highlighting the similarities with the optical filter. In Section 5, we apply both algorithms to galaxy catalogues from the COSMOS survey and we discuss the resulting cluster detections. We compare these results with
previous cluster catalogues obtained with different techniques on the same field.
Summary, conclusions and future prospects of this work are provided in
Section 6.

\section{Optimal filter for galaxy overdensities}

To derive the optimal filter, we assume that positions, magnitudes (at least in one band), and possibly photometric redshifts of the galaxies in a
surveyed field are available. Furthermore, we assume that bound structures like galaxy
clusters present in the field can be traced by the galaxy distributions within
them and that the spatial distribution of background galaxies is random. Under
these assumptions, given
the characteristics of the survey, bands of observation and depth,
it is possible to define a cluster model, describing the expected
spatial and magnitude distributions of the cluster galaxies and a noise model describing the background distribution of field galaxies.

Thus we define a model $n_m(\bmath \theta,m)$ for the observed distribution of galaxies as a function of the position $\bmath \theta$ and magnitude $m$, given by the sum of a
field and a cluster component \citep{b12}:
\begin{equation}\label{model}
n_m(\bmath \theta, m) = n_f(m) + \Lambda n_c(\bmath \theta, m) = n_f(m) + \Lambda
P(\bmath \theta - \bmath{\theta_c}) \phi(m)\;,
\end{equation}
where $n_f(m)$ is the magnitude distribution of field galaxies, $\bmath
\theta_c$ is the cluster centre, $P$ is the
projected radial profile of the cluster galaxies number density, $\phi(m)$ is the cluster
luminosity function and $\Lambda$ is a richness parameter, proportional to the total
number of cluster galaxies.

The detection of a cluster in the input catalogue is based on the
comparison of the observed distribution of galaxies to the model $n_m$. More
specifically, for a fixed cluster centre $\bmath {\theta_c}$,
we compute the value of $\Lambda$ for which our model distribution
best describes the observed distribution $n_d(\bmath \theta, m)$. The likelihood of the observed data $n_d$ given the model $n_m$ is
\begin{eqnarray} \label{like}
\mathcal L &=& - \int \frac {[n_d(\bmath \theta, m) - n_m(\bmath \theta,
m)]^2}{n_m(\bmath \theta,m)}d\Omega
dm \nonumber \\
&=& - \int \frac {[n_d(\bmath \theta, m) - n_f(m) - \Lambda n_c(\bmath \theta,
m)]^2}{n_f(m)}d\Omega
dm\;,
\end{eqnarray}
assuming that the background noise is Poissonian noise and that the galaxy density is high enough such that the Gaussian
approximation holds.

We derive the
value of the richness $\Lambda$ that maximizes $\mathcal L$ by imposing
\begin{equation}
\frac{d \mathcal L}{d\Lambda} = 2\int \frac{n_c}{n_f} (n_d-n_f) d\Omega dm -2\Lambda \int
\frac{n_c^2}{n_f} d\Omega dm = 0\;,
\end{equation}
which leads to
\begin{eqnarray} \label{lambda}
\Lambda &=& \frac {\displaystyle \int \frac {n_c}{n_f} (n_d - n_f) d\Omega
dm}{\displaystyle \int \frac {n_c^2}{n_f} d\Omega dm} \nonumber \\
&=& \frac {\displaystyle \int \frac {n_c}{n_f} n_d  d\Omega
dm - \int n_c d\Omega dm}{\displaystyle \int \frac {n_c^2}{n_f} d\Omega dm}\;.
\end{eqnarray}
Only the first term in the numerator depends on the spatial
distribution of the data with respect to $\bmath{\theta_c}$. Thus, Equation (\ref{lambda}) can be re-written
as
\begin{equation} \label{lambda2}
\Lambda = \int \Phi(\bmath \theta - \bmath{\theta_c}, m) n_d(\bmath \theta, m) d\Omega dm - B\;,
\end{equation}
where $\Phi$ is the optical linear filter defined as
\begin{equation}\label{filter}
\Phi (\bmath \theta - \bmath{\theta_c}, m) = \biggl( \displaystyle \int \frac {n_c^2}{n_f} d\Omega dm \biggr)^{-1} {\displaystyle \frac {n_c(\bmath \theta, m)}{n_f(m)}}\;,
\end{equation}
and $B$ is the contribution of background galaxies that is subtracted to work on a zero-mean noise field,
\begin{equation}
B =  \biggl( \displaystyle \int \frac {n_c^2}{n_f} d\Omega dm \biggr)^{-1} \int n_c d\Omega dm\;.
\end{equation}
If we insert $\Lambda$ obtained from Equation (\ref{lambda}) back into
Equation (\ref{like}), we obtain the corresponding value of the likelihood
\begin{equation}\label{L}
\mathcal L = \mathcal L_0 + \frac {\displaystyle \biggl[\int \frac {n_c}{n_f} (n_d - n_f) d\Omega
dm\biggr]^2}{\displaystyle \int \frac {n_c^2}{n_f} d\Omega dm}\;,
\end{equation}
where $\mathcal L_0$ is a (negative) constant that does not depend on the position of the cluster $\bmath \theta_c$. Comparing Equations (\ref{L}) and (\ref{lambda}), we see that the varying part of $\mathcal L$ is proportional to $\Lambda^2$. This squared dependence of $\mathcal L$ on $\Lambda$ reflects the fact that the $\chi^2$ approach returns high likelihoods also for galaxy underdensities, for which $\Lambda$ would be negative.

This is why we consider convenient to search for galaxy clusters as
peaks of the distribution of $\Lambda$ as a function of the sky position $\bmath{\theta_c}$, instead of likelihood maxima. In fact, we take full advantage of the linear response to the data, provided by Equation (\ref{lambda2}), and of the physical direct interpretation of $\Lambda$, which is a suitably normalized number of visible galaxies belonging to the cluster (see Section \ref{prac}).

The variance of $\Lambda$ is given by $\sigma^2_\Lambda \equiv \langle (\Lambda -
\hat \Lambda)^2 \rangle$, where $\hat \Lambda$ is an estimate of the true value of $\Lambda$, given by Equation (\ref{lambda2}). Since $\hat {\Lambda n_c}(m,\bmath \theta)$ and $\hat n_f(m)$ are random
realizations of
the cluster and field populations, the resulting $\sigma^2_\Lambda$ has two contributions: the first term is
induced
by the random fluctuations in the background, the second by the random
sampling of the cluster model:
\begin{equation}\label{variance}
\sigma^2_\Lambda = \left(\int \frac {n_c^2}{n_f} d\Omega dm \right)^{-1}+
\Lambda  \frac {\displaystyle \int \frac {n_c^3}{n_f^2}d\Omega
dm}{\displaystyle \left(\int \frac {n_c^2}{n_f}d\Omega dm \right)^2}\;.
\end{equation}
By applying the formalism of linear matched filters, one can see that the filter minimizes the variance of the estimated $\Lambda$  due to random fluctuations only. The second term in Equation (\ref{variance}) gives the intrinsic Poissonian fluctuation of the cluster realization.

\subsection{Modelling the galaxy clusters}\label{prac}

\begin{figure}
\begin{centering}
\includegraphics[scale = 0.65]{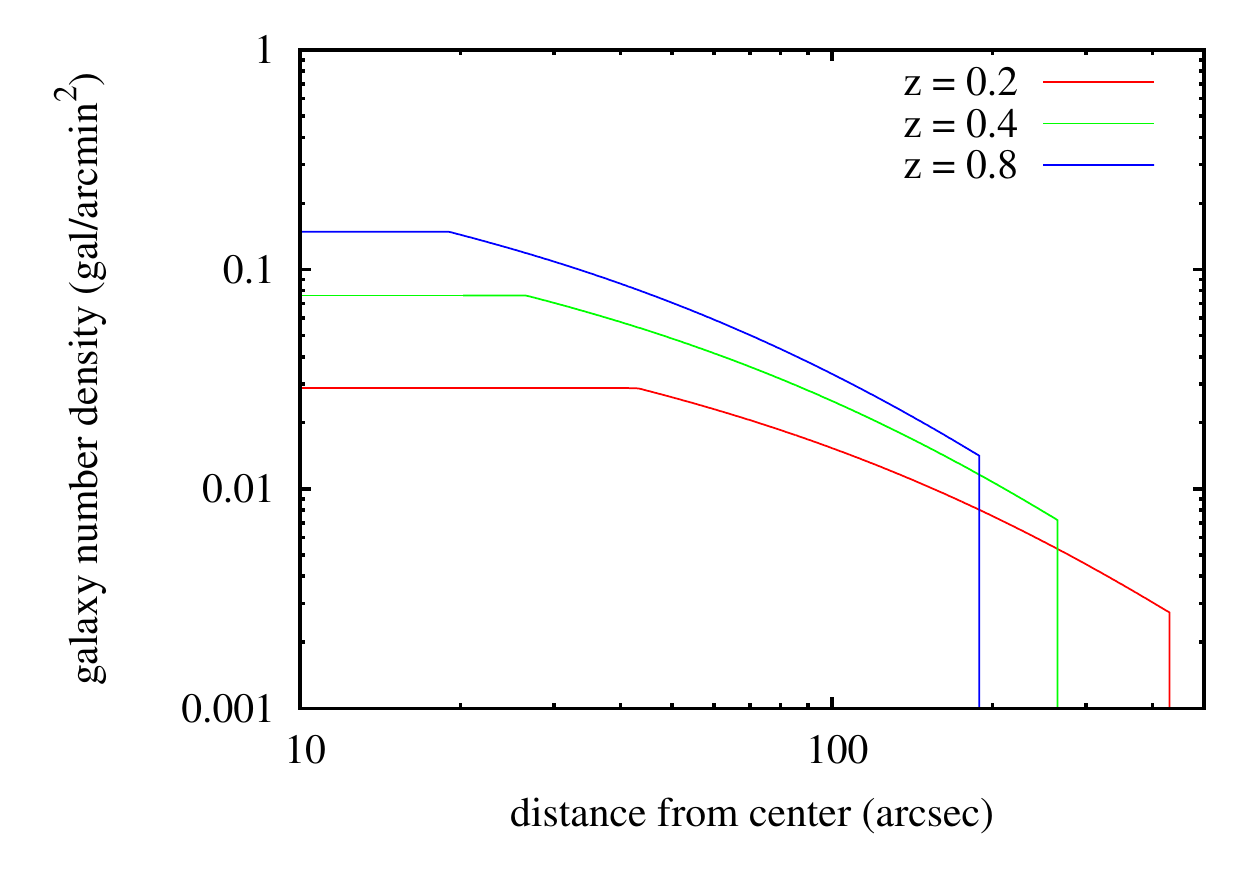}
\end{centering}
\caption{Radial profiles of the cluster model at different redshifts, as labelled in the figure. \textbf{The profile is flattened inside 100 Kpc/$h$ to avoid the central divergence, while the cut off at large radius corresponds to the $R_{200}$ limit.} }
\label{radProf}
\end{figure}

The cluster optical model is specified by the
spatial and luminosity distribution of cluster galaxies. For simplicity, we assume spherical symmetry. We also assume that the projected radial distribution of the cluster members follows an NFW model
\citep{b16,b17,b18}:
\begin{equation}\label{2dnfw}
\Sigma(x) = \frac{2 \rho_s r_s}{x^2-1} f(x),
\end{equation}
with
\begin{displaymath}
f(x) = \left\{ \begin{array}{ll}
1-\frac{2}{\sqrt {x^2-1}} \,  \textrm{arctan} \,   \sqrt{\frac{x-1}{x+1}} &
(x>1)\\
1-\frac{2}{\sqrt {1-x^2}} \,  \textrm{arctanh} \,  \sqrt{\frac{1-x}{1+x}} &
(x<1)\\
0 & (x = 1)
\end{array} \right.\;,
\end{displaymath}
where $x \equiv r/r_s$. This distribution, apart from the normalization $\rho_s$,
depends on the scale radius $r_s$ only, corresponding to $R_{200}/c$, where $R_{200}$ is the scale where the galaxy density is 200
times the critical density, and $c$ is the so-called concentration parameter. Our choice is
motivated by the analysis made by \citet{b3} on a large sample of optically-selected clusters from the SDSS. They found that the NFW model is a good
description of the cluster galaxy distribution up $R_{200}$. The
concentration parameters
derived from galaxy distributions described by the NFW model depends on the richness but it is
in general smaller ($1<c<3$) than those found in numerical simulations for dark matter halos of similar mass. In this context we use a scale radius $r_s$ = 500 kpc/$h$, that represents a good estimate
for quite rich clusters \citep[see Figs. 7 and 8 in][]{b3}.
The resulting radial profiles computed at different redshifts are shown in Fig. \ref{radProf}, as examples.

For the cluster luminosity distribution, we assume the Schechter luminosity functions \citep{b23}, with parameters taken from
\citet{b11}, who analysed 97 clusters observed in the SDSS. Since these are calculated from galaxy counts
inside a radius of $1$ Mpc/$h$ from the cluster centre, we apply the same cut to the radial density profile. For a realistic
concentration parameter $c$ = 2, this implies that we limit our analysis within $R_{200}$, where the projected NFW is a good
description of observed clusters, as discussed.

\begin{figure}
\begin{centering}
\includegraphics[scale = 0.65]{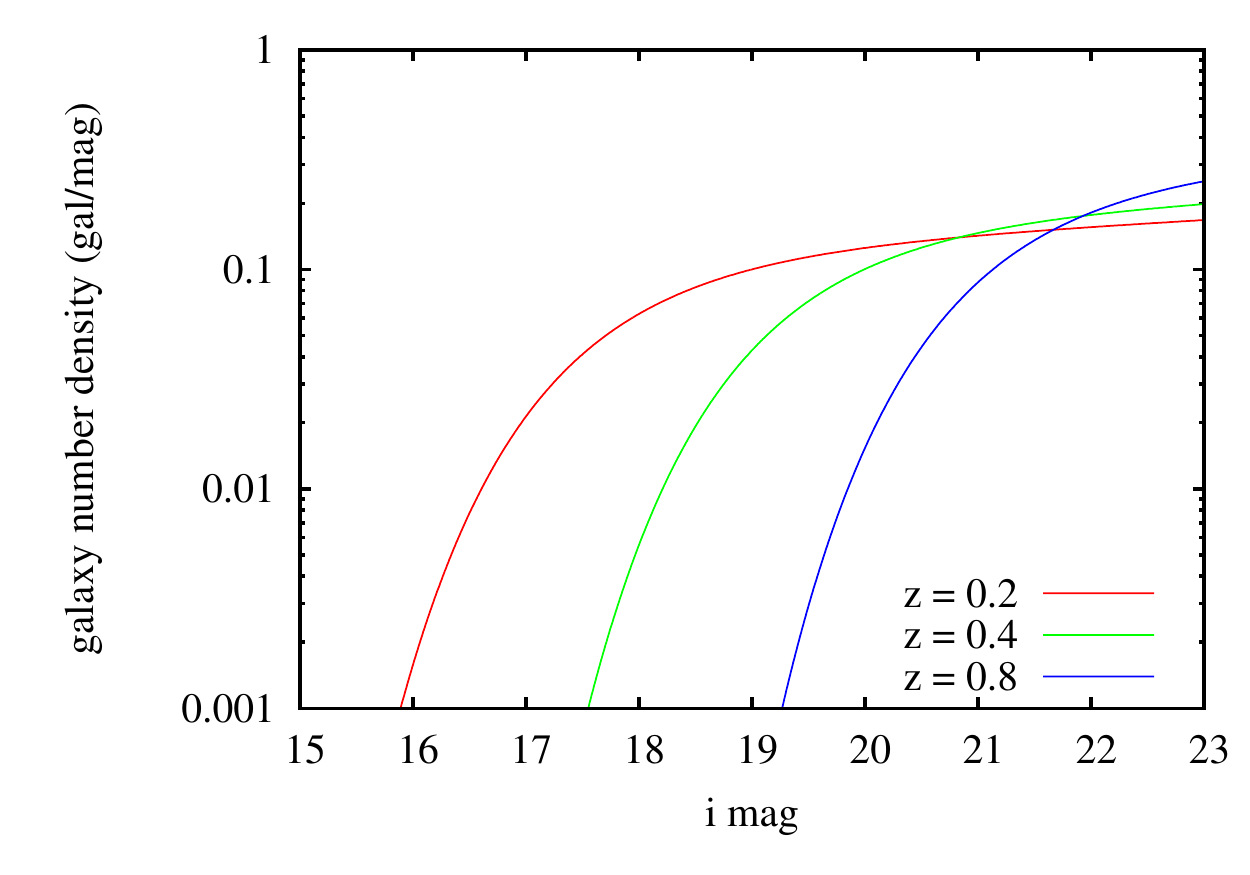}
\end{centering}
\caption{Magnitude functions of the cluster model at different
redshifts, as labelled in the figure.}
\label{magFilter}
\end{figure}

Apart from the cluster model, the other ingredient of our recipe is the
distribution of field galaxies, that we assume to be random with a
fixed mean angular number density. The magnitude distribution of the field galaxies
$n_f(m)$ can be
estimated from the distribution of the whole galaxy sample, provided that the
field is large
enough such that the contribution of cluster galaxies can be considered negligible.

\subsection{Map making}\label{mapmaking}

Clusters are searched in three-dimensional space, where their positions are given by the angular coordinates $\bmath{\theta_c}$ and by the redshift $z_c$. We create maps of $\Lambda$ estimates for different values of the redshift, and we search for cluster detections in these maps. Comparison of detections in different redshift slices is done subsequently (see Section \ref{nored}).

For every redshift, we adapt the cluster model $n_c$ in a proper way. First, we transform the cluster absolute luminosity function into an apparent magnitude distribution (see Fig. \ref{magFilter}), and re-scale the radial profile of the cluster model according to the angular-diameter distance. Then, we normalize properly the cluster model $n_c$ to set the richness parameter $\Lambda$ to be the total number $N$ of visible galaxies in a cluster. Integrating the cluster model, we get:
\begin{eqnarray}
N &=& \Lambda \int n_c(\bmath  \theta,m) d^2\theta dm \\
  &=& \Lambda \int \!\!\! \int P(\bmath \theta) d^2 \theta \int \phi(m) dm \;.
\end{eqnarray}
To set $N = \Lambda$, we impose the integrals of the angular distribution and the luminosity function to be
normalized to unity, that is
\begin{equation}
\int \!\!\! \int P(\bmath \theta) d^2 \theta = 1\;,
\end{equation}
\begin{equation}
\int_0^{m_{\rm lim}} \phi(m) dm = 1\;,
\end{equation}
where $m_{\rm lim}$ is the limiting magnitude of the sample.

At a fixed redshift $z_c$, we build a two-dimensional map evaluating $\Lambda$ over a grid
of angular positions $\bmath{\theta_c}$. Since we
deal with discrete quantities, the integral in Equation (\ref{lambda}) must be approximated as a sum over the galaxy positions. Only the first term in the
numerator needs to be evaluated for every grid position
$\bmath{\theta_c}$, while the second one is equal to unity because of the normalization. The denominator for a given redshift is a constant for all sky positions. We
denote it by $C$. Thus the discretized version of Equation (\ref{lambda}) reduces to
\begin{equation} \label{pract}
\Lambda_{ij} = \Lambda(\theta_i, \theta_j) = \frac {A_{ij}-1}{C}\;,
\end{equation}
where
\begin{equation} \label{practnum}
A_{ij} = \sum_{k=1}^{N} \frac {n_c(r_k, m_k)}{n_f(m_k)}= \sum_{k=1}^{N} \frac {P(r_k) \phi(m_k)}{n_f(m_k)}\;.
\end{equation}
In the formula above, $r_k$ is the angular distance of the galaxy from the centre
\begin{equation}
r_k = |\bmath{\theta_k} - (\theta_i,\theta_j)|\;,
\end{equation}
and the index $k$ runs over all $N$ galaxies of the sample. Clusters candidates are identified with the peaks of the resulting $\Lambda$-map. To select only significant peaks, we calculate their signal-to-noise ratios using the variance given in Equation (\ref{variance}).

\subsection{Redshift determination}\label{reddet}
\subsubsection{Without photometric redshifts}\label{nored}

\begin{figure}
\begin{centering}
\includegraphics[scale = 0.65]{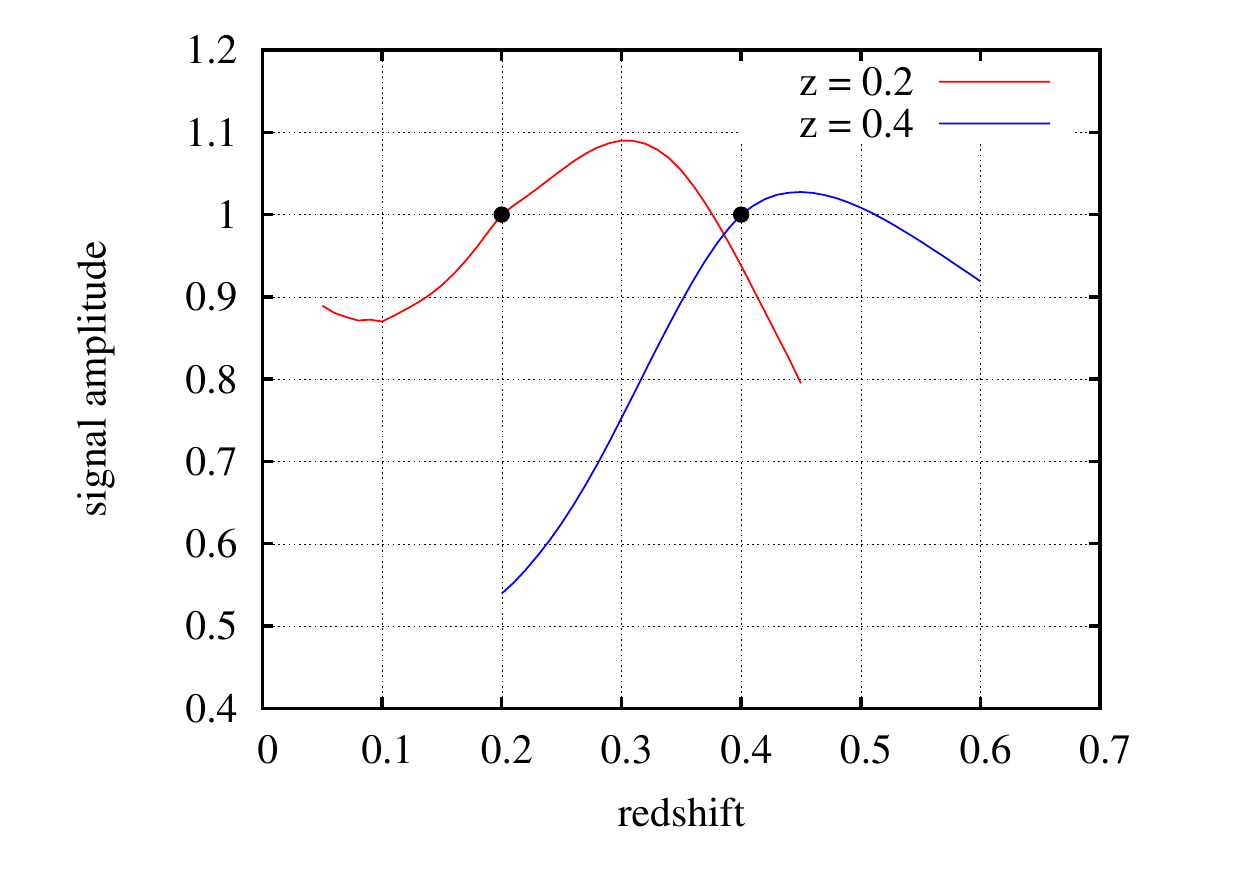}
\end{centering}
\caption{Analytic response of the algorithm (Eq. \ref{anresp}) for two clusters at redshifts 0.2
and 0.4. When the cluster is seen at the correct redshift, the value of the
response is unity by definition (black points).}
\label{rbias}
\end{figure}

To estimate the cluster redshift, we cannot just maximize $\Lambda$ with respect to $z_c$. Indeed, this would introduce a bias as discussed below. The analytic response of
the algorithm as a function of the search redshift $z_c$ for a cluster at a redshift $\bar z_c$ and with a galaxy distribution $\bar n_c$ is

\begin{equation}\label{anresp}
\Lambda(z_c) = \frac{\displaystyle \int \frac{n_c}{n_f} \bar n_c d \Omega
dm}{\displaystyle \int \frac{n^2_c}{n_f} d \Omega dm}\;.
\end{equation}
This corresponds to Equation (\ref{lambda}) when applied to a known distribution $n_d =
n_f+\bar n_c$. By construction, $\Lambda = 1$ when $z_c = \bar z_c$, but
there is no reason why this should be a maximum for the
function $\Lambda(z_c)$. This is shown in Fig. \ref{rbias}, where we plot the
analytic response for two clusters located at $z = 0.2$ and $z = 0.4$. For
both of them the maximum of $\Lambda$ as a function of $z_c$ is found at a
higher redshift than the cluster's.

\begin{figure}%[htb]
\begin{centering}
\includegraphics[scale = 0.65]{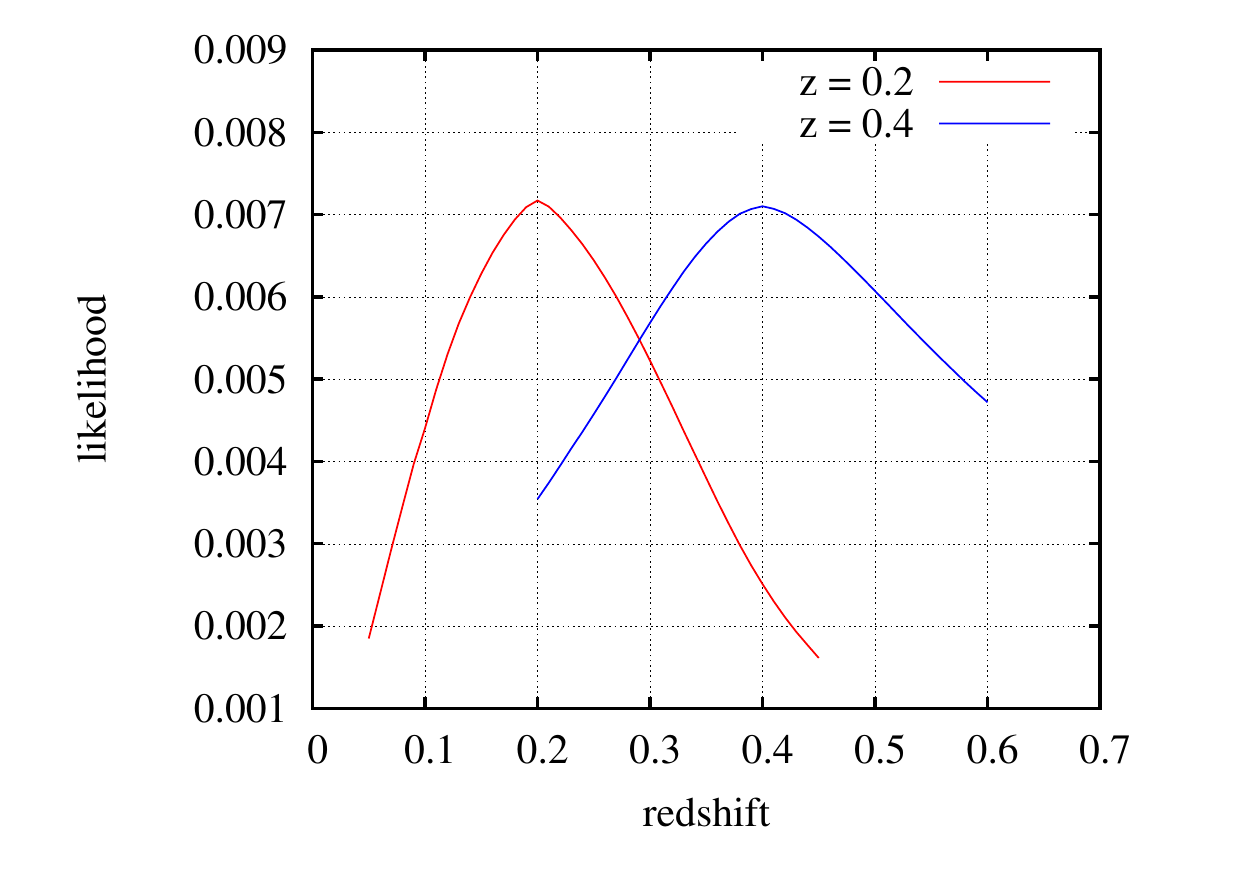}
\end{centering}
\caption{Analytic value of $\mathcal L$ for two clusters at redshift 0.2 and
0.4, as a function of the search redshift $z_c$.}
\label{unbias}
\end{figure}

For a correct redshift estimate, the likelihood $\mathcal L$ (Equation
\ref{L}) in
the peaks of the 2D distribution of $\Lambda$ must be used. By
definition, the
value of $\mathcal L$ is maximal when the cluster is filtered by a model at the correct
redshift (See Fig. \ref{unbias}).

\subsubsection{With photometric redshifts}\label{photoz}

If the photometric redshift is available for each galaxy, we can use this further information in our analysis. Assuming that a redshift
probability distribution $p_k(z)$ is derived for the $k$-th galaxy of the sample, $p_k(z_c)$ can be used as a weight factor for that galaxy when estimating the field number density, $n_f$, the cluster richness, $\Lambda$, and the likelihood, $\mathcal L$, at a redshift $z_c$..

On the other hand, this can be interpreted as an
extension of the two-dimensional linear matched filter to a three-dimensional one, where the redshift probability function represents the cluster profile along the line of sight. In fact, in an ideal case where all galaxies have the same redshift probability distribution, which for simplicity we assume to be a Gaussian distribution centered on their most probable location, $z_{est,k}$, with a constant rms, $\sigma_z$,
\begin{equation}
p_k(z) = \frac 1 {\sqrt {2\pi} \sigma_z} \, \exp \, \bigg(- \frac {(z_{est,k}
- z)^2}{2\sigma_z^2}\bigg)\;,
\end{equation}
the three-dimensional model with a cluster at redshift $z_c$ is
\begin{eqnarray}\label{model_z}
n_m(\bmath \theta, m, z) &=& n_f(m,z) + \Lambda n_c(\bmath \theta, m, z) \nonumber
\\
&=& n_f(m,z) + \Lambda P(\bmath \theta - \bmath{\theta_c}) \phi_z(m) p(z)\;,
\end{eqnarray}
where
\begin{equation}
p(z) = \frac 1 {\sqrt {2\pi} \sigma_z} \, \exp \, \bigg(- \frac {(z_c
- z)^2}{2\sigma_z^2}\bigg)\;.
\end{equation}
If we then further assume that the background distribution is only slowly changing with redshift, so that the background $n_f(m,z)$ for $z_c - \sigma_z < z < z_c+\sigma_z$ is constant,
the richness estimate at redshift $z_c$ is
\begin{equation} \label{lambda_z}
\Lambda = \frac {\displaystyle \int \frac {n_c(\bmath
\theta,m,z)}{n_f(m,z_c)}
(n_d(\bmath \theta,m,z) - n_f(m,z_c)) d\Omega
dm dz}{\displaystyle \int \frac {n_c^2(\bmath \theta,m,z)}{n_f(m,z_c)}
d\Omega dm dz}\;.
\end{equation}
This is equivalent to Equation (\ref{lambda}), but with the additional dimension. In practical applications, Equation (\ref{lambda_z}) has to be discretized in analogy with Equation (\ref{pract}). Thus,
\begin{equation}
A_{ij} = \sum_{k=1}^{N} \frac {P(r_k) \phi(m_k)p_k(z_c)}{n_f(m_k,z_c)}\;.
\end{equation}
In this case the numerator is evaluated weighting every galaxy with $p_k(z_c)$. The constant denominator $C$, that does not depend on the observed galaxy spatial distribution, can be evaluated using a model for $\sigma_z$, possibly depending on the magnitude
and the redshift estimate of the galaxies. The same model is also necessary for the
evaluation of the variance that, in analogy with Equation \ref{variance}, is
\begin{equation}\label{variance_z}
\sigma^2_\Lambda = \left(\int \frac {n_c^2}{n_f} d\Omega dm dz \right)^{-1}+
\Lambda  \frac {\displaystyle \int \frac {n_c^3}{n_f^2}d\Omega
dm dz}{\displaystyle \left(\int \frac {n_c^2}{n_f}d\Omega dm dz \right)^2}\;.
\end{equation}

In the process of estimating the redshift from multi-band photometry, a spectral type is also assigned to each galaxy. This can be used to k-correct the values of the magnitude and to obtain rest-frame values, which are directly comparable to the model. The filter is applied to the data as usual, creating maps for different redshift slices and selecting peaks, as described in Section (\ref{mapmaking}). The most likely redshift estimate for a detection is then choosen.

\subsection{Comparison with other optical cluster finding methods}

The optimal matched filter we used for cluster detections (both
optical and weak lensing) relies on the general knowledge we have of
galaxy clusters, as incorporated in the cluster model, and selects all
physical properties which enables us to distinguish them from the
field. It is in fact a general approach
which can incorporate at once different cluster aspects in contrast to other
methods which aim at a single signature such as galaxy overdensities,
cluster galaxy red-sequence, cD galaxies, etc. In other words,
instead of enforcing some criteria which is later applied to data, we
first define the general properties of galaxy clusters and then let
the filter find what are the distinctive features for their detection
given the actual data noise.

In the following sections we discuss the main techniques applied
 in literature and how they compare with the one presented in this
 work.

\subsubsection{Galaxy overdensities}

Looking for galaxy overdensities in astronomical images by eye was the
first method used to detect galaxy clusters. In the last decades
different automated methods were proposed, such as friend-of-friend
algorithms \citep{FoF}, Voronoi tesselation \citep{Voronoi}, filtering
by wavelets \citep{b2}, adapting kernels \citep{kernels} or matched
filters \citep[e.g.][]{b12,b5,gilbank,b1,mena,b19}. Together with the
additional features discussed in the following sections, our algorithm
incorporates the mentioned matched filters. In fact it includes
 the angular clustering information on the sky according to the $P$
 term of Equation (23) or, if photometric redshifts are available, the
 full three-dimensional clustering information thanks to the $p$ term of
 Equation (23). In more details, the main differences with respect to the matched filters
 mentioned in this section are:
\begin{enumerate}
\item our filter can take advantage of the information from
 photometric redshifts in a very flexible way, accounting for their uncertainty. Instead of simply slicing the data-set and
 neglecting all galaxies outside a fixed redshift range
\citep[e.g.][]{b19}, we consider all galaxies belonging to the
  data
 set by weighing them according to their own redshift probability
 distribution.
\item It uses distinct luminosity functions for galaxy clusters
  and field galaxies to define the cluster model and the noise
  contribution. Although this seems an obvious choice, 
 previous filtering algorithms use the luminosity functions
 taken from the whole population of galaxies, including the field,
 to define the cluster model.

\item It is
 linear with respect to the data, because it makes use of the
 richness $\Lambda$
 instead of the likelihood $\mathcal L$ as the main object of analysis.
 In fact, even if they are based on the same statistal distribution,
 $\Lambda$ is linear, in contrast with $\mathcal L$ which is
 quadratic, and therefore is capable to distinguish positive
 overdensities from negative underdensities to which $\mathcal L$
 would assign high probabilities. 
 
\item It gives as a
 natural output an estimate of the number of observed galaxies in a
 cluster, which can be easily corrected for redshift dimming to get a
 meaningful physical quantity (see Section \ref{optcosm}).

\end{enumerate}

\subsubsection{Brighter cluster galaxy}\label{cd}

The algorithm used by \cite{maxbcg}, based mainly on
red-sequence information, includes a cut-off term for the brighter cluster galaxy luminosity, such that detections without a very brilliant central galaxy are discarded.
In our approach, instead of looking for a
 specific cluster member, we let the filter select in an optimal way the whole
 luminosity function $\phi$ expected for a cluster at a given redshift in
 contrast with the field luminosity function $n_f$, as in Equation (23).
 As the most massive and luminous galaxies are found preferably at
 the center of the clusters, at low magnitudes the ratio between the
 cluster model luminosity function and the observed field
 distribution will increase.
 The algorithm will assign large weights to very luminous galaxies and thus they will be a strong indication
 of the presence of clusters.
 The advantage of our approach is that the filter is defined
 according to well defined statistical quantities and not to a free
 parameter. In addition we allow the algorihm to be
 sensitive to a broad range of systems, from rich clusters to groups
 which might otherwise fall below a given threshold.

\subsubsection{Cluster galaxies red sequence}

Red-sequence cluster finders rely on the fact that the center
of galaxy clusters is typically populated by elliptical galaxies whose
color is on average redder than the one of the field galaxies at the
same redshift. Usually a cut in a colour-magnitude diagram is
performed to select these red galaxies overdensities that identify the
galaxy cluster candidates.
This technique, presented in \citet{g&y}, has been succesfully used in
different galaxy surveys \citep[e.g.][]{rsc,maxbcg,lu,than}. We
 thus mention and explain how this technique can be implemented in
 the framework of our method with a simple generalization, even if we
 did not explicitly made use of color information in this work. In a
 following paper we will fully discuss a multi-band
 approach which deserves an extended discussion on its own.

 Our method belongs to the matched filters developed for CMB
 observations \citep[e.g.][]{sch,b10} which
 use at once all available bands to obtain a unique response. With
 this respect, the algorithm presented here is a special case where a single band is considered,
 but could be naturally extended to include the full available color
 information. In this way, we would consider not only the two bands that are usually
 used in classical red-sequence searches,
 but all bands and color properties of the clusters and
 the field, adding new possible information.
 Again, this would be done not by defining a cut-off in the color
 space but by defining the proper weight to each galaxy performing an
 optimized analysis.

 In any case, already at this stage without multi-band
 information, we partially account for the massive
 galaxies, as those targeted more precisely by the red sequence,
 thanks to the use of the cluster luminosity function (see Section
 \ref{cd}).

\section{Investigating the filter properties}\label{tests}
The final step of this work will be the application of the algorithm to real data, namely the $i'$ band galaxy catalogue of the COSMOS field \citep{b6}, limited at $i'$ = 25. Before doing this, we have to evaluate the capability of our algorithm to detect clusters and to correctly measure their richness and redshift. In these tests we use the field distributions measured from the COSMOS data to define the filter. Thus, we use the specific implementation of the algorithm that will be later adopted for the final data analysis.

\subsection{Signal-to-noise estimates}

\begin{figure}
\begin{centering}
\includegraphics[scale=0.7]{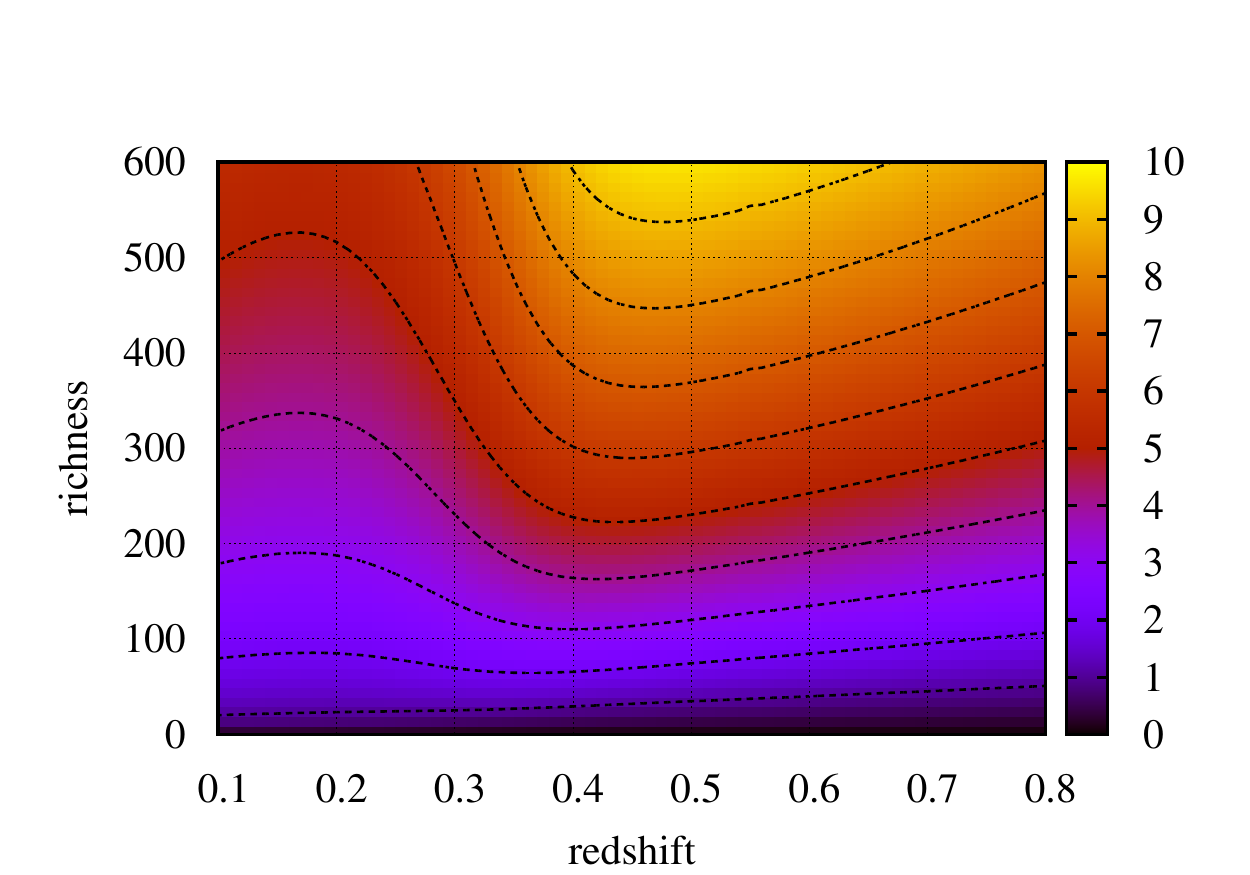}
\includegraphics[scale=0.7]{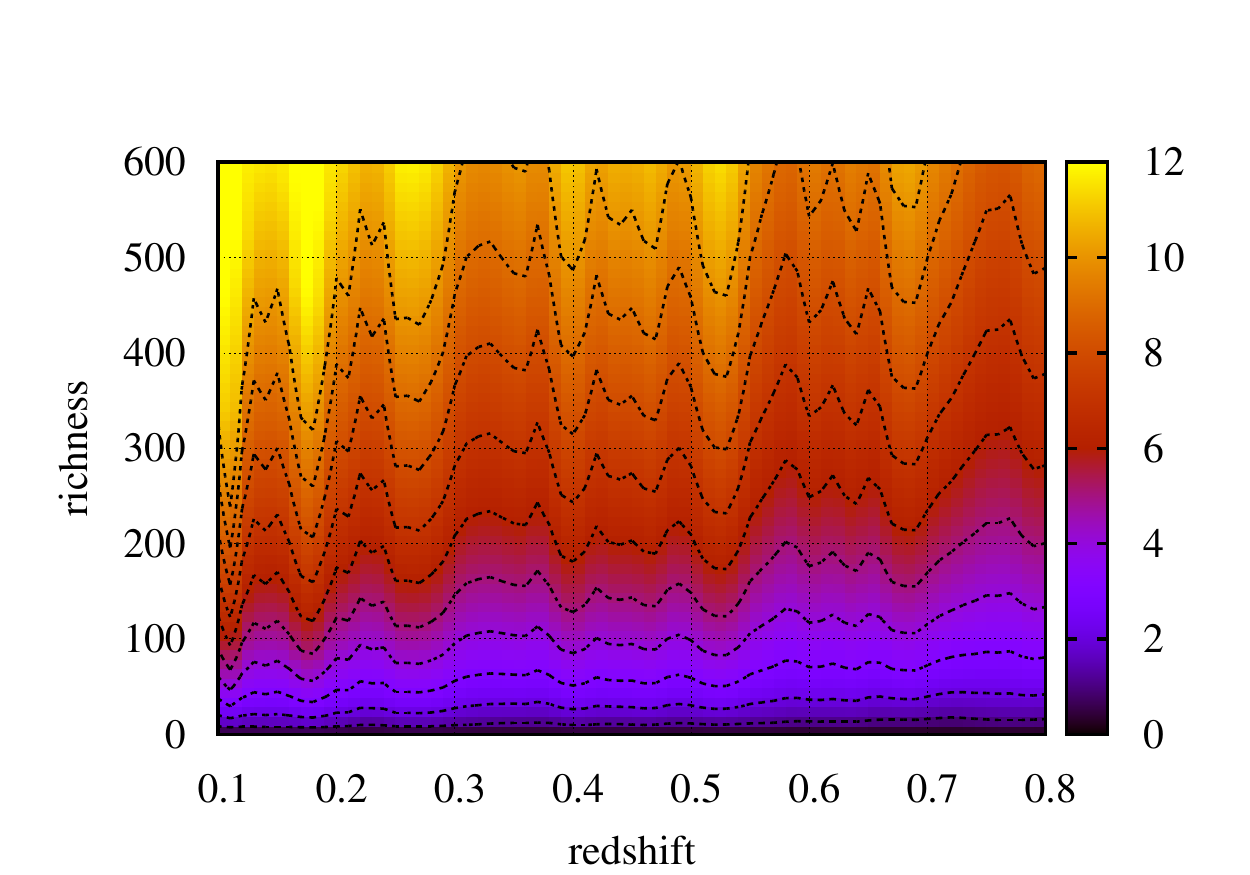}
\end{centering}
\caption{Expected signal-to-noise values for the detection of a cluster, without
(top panel) and
with (bottom panel) photometric redshifts, for the COSMOS $i'$ band catalogue. The black contours refer to different integer values for S/N, starting from unity at the bottom (see the color bar on the right). The value on the y-axis is the number of galaxies of the cluster that are detected when it is located at $z = 0.2$.}
\label{sn}
\end{figure}

\begin{figure}
\begin{centering}
\includegraphics[scale=0.7]{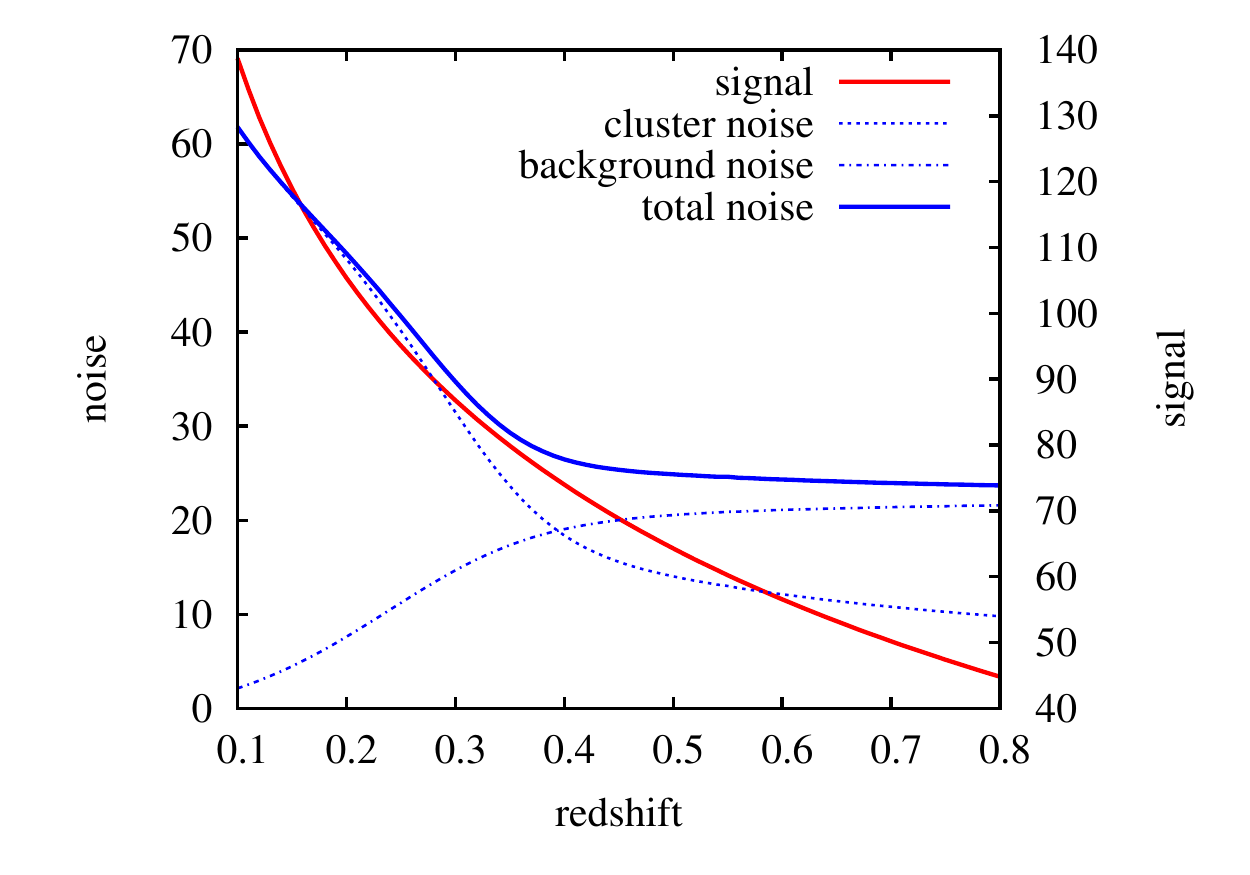}
\end{centering}
\caption{Signal and noise for a cluster with 100 visible galaxies at $z = 0.2$ as a function of redshift. The signal (in red) corresponds to the number of galaxies below $i' = 25$, calculated from the model luminosity function. The dot-dashed line represents the noise due to the background galaxies, while the dashed line represents the noise due to the fluctuations in the cluster galaxy distribution. The solid blue line represents the total noise. Scales for noise and signal are shown on the left and right, respectively.}
\label{s&n}
\end{figure}

We now evaluate the expected signal-to-noise ratios for clusters with different redshift and richness, as described by our cluster model, once the actual properties of the galaxies in the COSMOS
catalogue are assumed.
The expected signal-to-noise ratios when the photometric redshifts are neglected are shown in the upper panel of Fig. \ref{sn}. In this case,
the expected signal and noise are given by Equations (\ref{lambda}) and
(\ref{variance}), respectively, and the signal-to-noise ratio peaks at
$z \sim 0.45$. To explain this behaviour, we plot separately in Fig. \ref{s&n} the expected signal of a cluster with 100 visible galaxies at $z = 0.2$ and the two noise terms as a function of redshift. The signal, i.e. the number of visible
galaxies of the modeled cluster, decreases for increasing $z$ as
expected (solid line). The trend for the noise is different: on one hand, the term related to
the background galaxies only (dot-dashed line) grows monotonically
as a function of redshift up to $z\sim 0.5$ after which it basically
stays constant;
on the other hand, the intrinsic fluctuations of the cluster signal
decreases monotonically with $z$. As a result the latter term is dominant at low redshifts (up to $z \lesssim 0.4$).

This noise behaviour derives from the fact that for low-redshift
clusters, the filter, being proportional to $n_c(m)/n_f(m)$, selects the
bright end of the cluster luminosity function, that has almost no
background. Therefore the background noise decreases with redshift
while the intrinsic fluctuations of the cluster signal increase
because the $\Lambda$ estimate depends in practice on the very few
bright galaxies only.

In order to consider the case in which photometric redshift information is available, we use the values for the redshift ($z_{est}$) and its 68\% confidence levels
($z_{min}$, $z_{max}$) reported in the COSMOS catalogue to associate to each
galaxy a probability distribution given by
\begin{equation} \label{gauss}
p(z) =
  \frac 1 {\sqrt {2\pi} \sigma_m} \, \exp \, \bigg(- \frac {(z -
    z_{est})^2}{2\sigma^2}\bigg)\;,
\end{equation}
where
\begin{equation}
  \sigma = \left\{ \begin{array}{ll}
      z_{est} - z_{min} & \textrm{if $z < z_{est}$}\\
      z_{max} - z_{est} & \textrm{if $z > z_{est}$} \end{array}
  \right. \;,
\end{equation}
and
\begin{equation}
  \sigma_m = (z_{max} - z_{min})/2\;.
\end{equation}

When estimating the normalization constants, i.e. the
denominator of Equation (\ref{lambda_z}) and the variance (Equation \ref{variance_z}), which do not depend on the observed spatial galaxy distribution, we use the
mean redshift error for different classes of objects, estimated from
the comparison with a spectroscopic subsample \citep{b6}. For our dataset, these errors are:
\begin{equation}\label{sigma_z}
\sigma_{{\Delta z}/{1 + z}} =
\left\{
\begin{array}{lll}
0.007 & \textrm{for $0.2 < z < 1.5$,} & {i' < 22.5}\\
0.011 & \textrm{for $0.2 < z < 1.5$,} & {22.5 < i' < 24}\\
0.053 & \textrm{for $0.2 < z < 1.5$,} & {24 < i' < 25}\\
0.06 & \textrm{for $z > 1.5$}
\end{array} \right.\;.
\end{equation}

As we can see in the lower panel of Fig. \ref{sn}, using the photo-$z$ information, the picture becomes more confused,
as the sensitivity depends on the field population density
$n_f(m,z)$ at the search redshift (see Equation \ref{variance_z}).
The only clear trend is the decrement of the sensitivity at high
redshift due to the dimming of the cluster signal. In this case we do not have any worsening of the performance at low redshift because
the weight given by the filter through the $n_c(m)/n_f(m,z)$ ratio
selects almost the same part of the luminosity function at each
redshift.

\subsection{Simulations with mock catalogues}\label{mock}

\begin{figure}
\begin{centering}
\includegraphics[scale = 0.75]{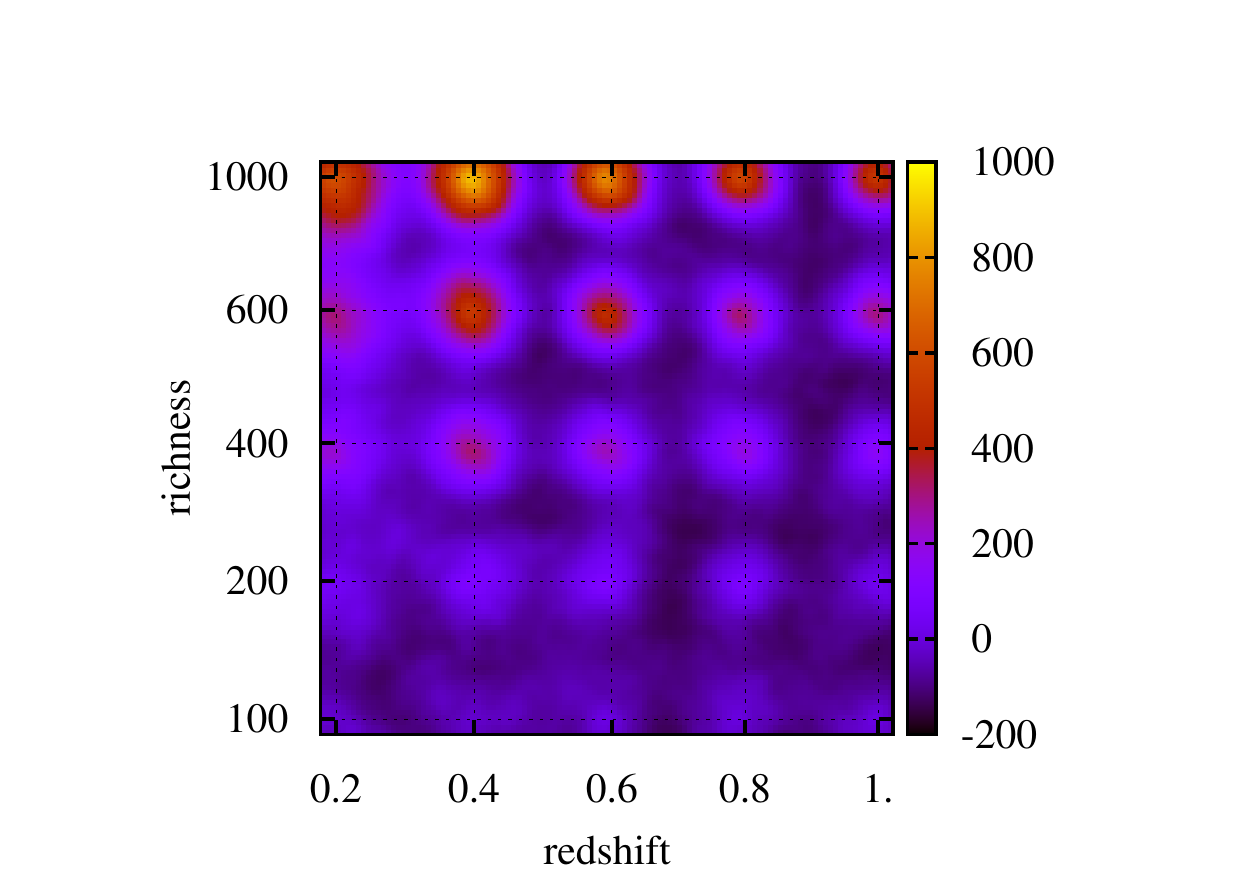}
\end{centering}
\caption{Distribution of the richness $\Lambda$ at $z_c$ = 0.4. Colorbar for $\Lambda$ is reported on the right. The grid identifies the input
positions of the mock clusters. The labels along the axes indicate the richness and redshift of the cluster at that position.}
\label{map}
\end{figure}

We then test the application of our algorithm to mock galaxy catalogues, built from the COSMOS data. To create realistic
catalogues for the field galaxies, we first randomize the positions of
the galaxies to cancel any structures contained in the field.
We assume that the cluster galaxies do not affect the photometric
properties of the overall sample because their number is negligible
with respect to that of the field galaxies. On the top of this random
background, we add some mock clusters. The cluster galaxies are
distributed according to the model for the spatial and luminosity
distribution of the cluster members (see Section \ref{prac}). The photometric
redshifts are assigned following a Gaussian redshift probability
distribution with errors given by Equation (\ref{sigma_z}). For illustrative purposes, we
first generate a square field with side 1 degree, containing 25
galaxy clusters of different richness and redshift placed on a
regular grid. In Fig. \ref{map} we show the results of the analysis of
this field without the usage of photometric redshifts, at a search redshift $z_c$ = 0.4. All clusters leave an
imprint on the $\Lambda$-map, whose strength depends on the richness
and the redshift, with clusters at $z=z_c$ being brighter than the
others with the same $\Lambda$.

\begin{figure}
\begin{centering}
\includegraphics[scale = 0.65]{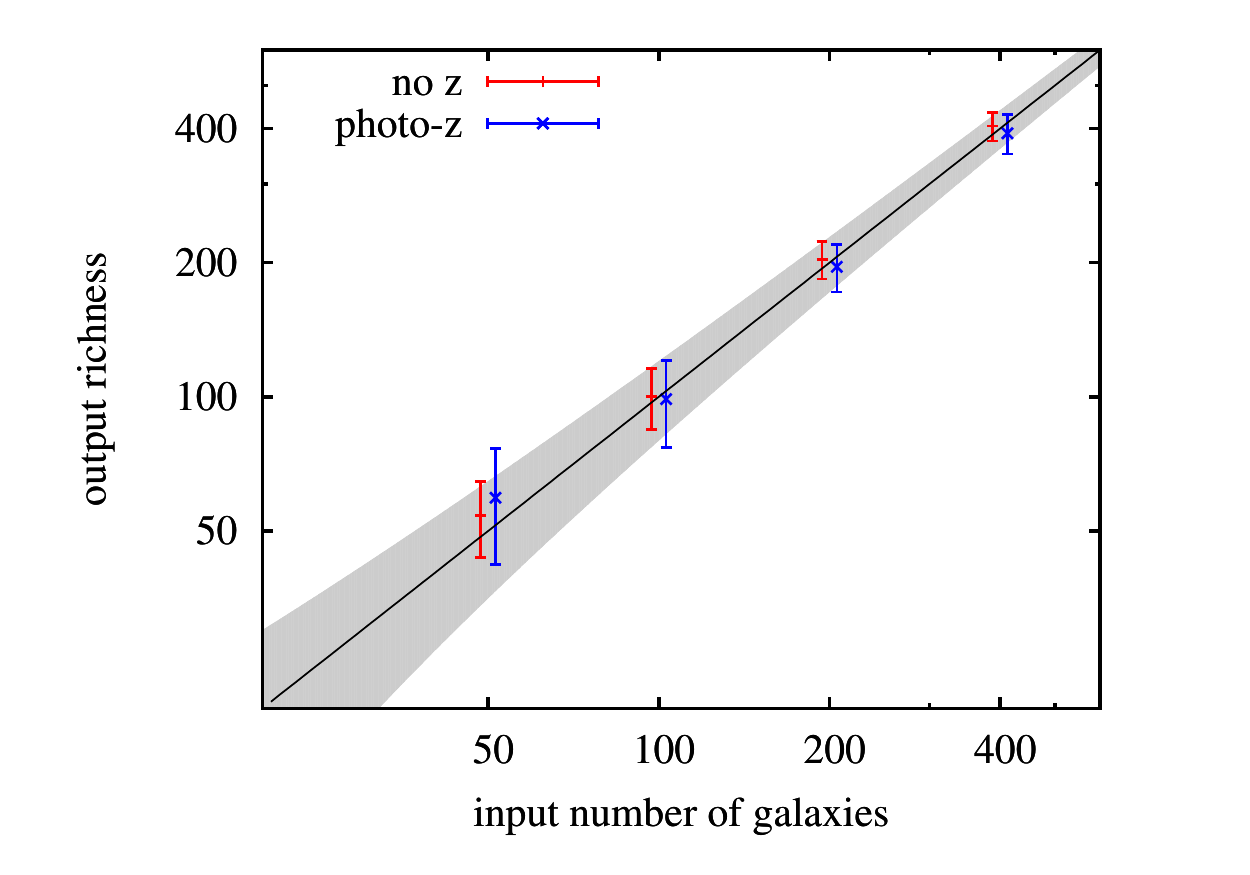}
\end{centering}
\caption{Calibration of the values for the richness $\Lambda$ values obtained by the algorithm with and
without redshift information. Clusters are located at $z = 0.4$. The shaded area represents the analytic estimate of the uncertainty, when the photometric redshift information is used.}
\label{norm}
\end{figure}

Other simulations are done to verify the linear response of the filter
with respect to the cluster richness. In this case, we assume for
simplicity all clusters at $z = 0.4$ and compare the input number of
galaxies to the estimated one. The results are shown in
Fig.~\ref{norm} where we also verify the good agreement of the
analytic estimate of the variance given in Equation (\ref{variance_z}) with
the one resulting from our Monte Carlo simulations.

\begin{figure}
\begin{centering}
\includegraphics[scale = 0.65]{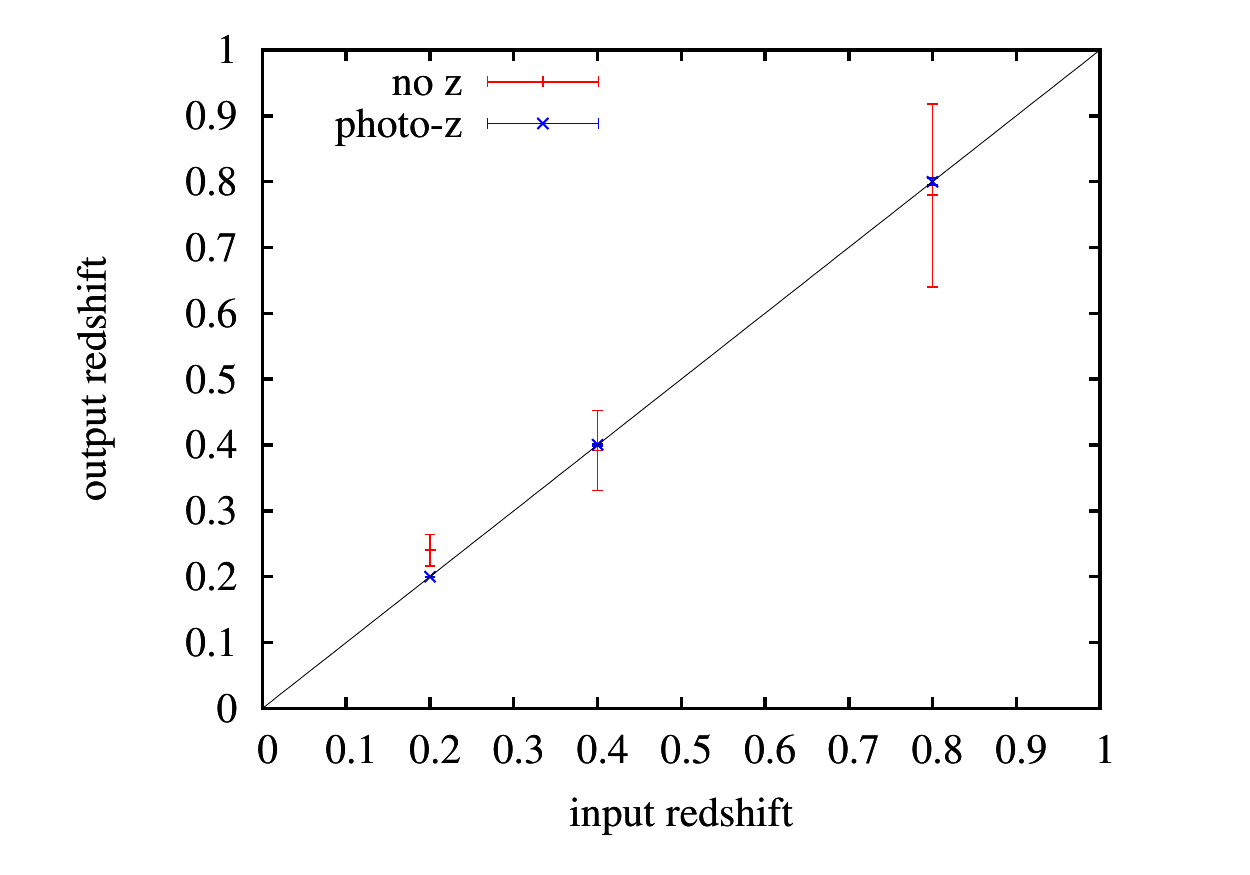}
\end{centering}
\caption{Redshift determination obtained by the algorithm, using the likelihood $\mathcal L$ instead of the richness $\Lambda$, with and without redshift information. When photometric redshift are used, error bars become negligible.}
\label{rnorm}
\end{figure}

We also use simulations to test the capability of the algorithm to estimate correctly the redshift of
a cluster. The results are shown in
Fig.~\ref{rnorm}. The mock clusters were composed
requiring that they have 200 galaxies under the magnitude limit at $z= 0.2$. At higher redshifts, fewer galaxies are seen and
analysed, explaining the increasing variance of the measurement. Note
that even with single-band observations a redshift estimate for the
clusters is possible albeit with larger errors with respect to the
case in which the photometric redshift information is included. A similar result was obtained by \citet{dietrich}.

\section{Optimal filter for weak lensing}\label{lenssect}

In the next Section we will apply to the COSMOS field both optimal filters for galaxy cluster detection from optical and weak lensing data. Here we introduce weak gravitational lensing and then the filtering method we will adopt for lensing data.

The observable quantity in weak lensing analyses is the reduced shear $g$, a measure of the distortion of background galaxies by intervening structures (i.e. galaxy clusters). From the 2D projection of the cluster mass distribution, $\Sigma (\bmath \theta$), one can calculate the lensing potential
\begin{equation}
\phi(\bmath \theta) = \frac{4G}{c^2} \frac{D_l D_s}{D_{ls}} \int d^2 \theta^\prime \Sigma (\bmath \theta^\prime) \ln|\bmath \theta - \bmath {\theta^\prime|}\;,
\end{equation}
where $D_l$, $D_s$ and $D_{ls}$ are the angular diameter distances between lens and observer, source and observer, and lens and source, respectively. 
The complex reduced shear is then defined as
\begin{equation}
g(\bmath \theta) \equiv \frac{\gamma(\bmath \theta)}{1 - \kappa(\bmath \theta)}\;,
\end{equation}
where the shear $\gamma$ and the convergence $\kappa$ are second-order derivatives of the lensing potential, namely
\begin{equation}
\kappa = \frac{1}{2}(\phi_{11} + \phi_{22})\;,\\
\end{equation}
\begin{equation}
\gamma_1 = \frac{1}{2}(\phi_{11} - \phi_{22})\;,\\
\end{equation}
\begin{equation}
\gamma_2 = \phi_{12}\;,\\
\end{equation}
using the notation
\begin{equation}
\phi_{ij} = \frac{\partial^2 \phi}{\partial \theta_i \partial \theta_j}\;.
\end{equation}

The linear matched filter for weak lensing cluster detections has been presented in \citet{b7}. We refer the reader to that paper and to \cite{b9} for a detailed description of its specific derivation, the comparison with other filtering techniques and tests of its performances. Here we summarise the main properties of the filter and how it can be applied to real data.

The weak lensing filter for cluster detection is expressed in Fourier space for convenience. Its shape is given by
\begin{equation} \label{lensfil}
\hat \Psi(\bmath k) = \displaystyle \frac 1{(2 \pi)^2} \biggl( \displaystyle \int \frac{\hat\tau^2(\bmath k)}{P_N(k)}d^2k \biggr)^{-1}{\displaystyle  \frac {\hat
\tau(\bmath k)}{P_N(k)}}\;,
\end{equation}
where $\hat \tau$ is the Fourier transform of the weak lensing signal expected
from a lensing cluster, i.e. the reduced shear, and $P_N$ represents the power spectrum of the noise, due to the intrinsic ellipticity of the sources, their finite number on the sky and the additional shear induced by large-scale structures. The weak lensing signal model for clusters is computed assuming a spherically symmetric NFW density profile for the dark matter distribution (Equation \ref{2dnfw}).
Again, as in the case of galaxy overdensities, the estimate of the lensing signal is obtained by convolving the shear data, $D(\bmath \theta)$, with the filter $\Psi$,
\begin{equation} \label{lenssign}
A(\bmath \theta) = \int {D(\bmath{\theta^\prime}) \Psi(\bmath \theta - \bmath{\theta^\prime}) d^2\theta^\prime}\;.
\end{equation}
The variance of the estimate is
\begin{equation}\label{sigma_A}
\sigma_A^2 = \frac{1}{(2\pi)^2} \int |\hat \Psi(\bmath k)|^2 P_N (k) d^2k\;.
\end{equation}
The filter in Equation (\ref{lensfil}) was constructed requiring that the estimate of the amplitude of the signal $A$ given by Equation (\ref{lenssign}) is unbiased and has minimum variance. The optical filter is the solution of the same problem for a galaxy distribution in magnitude and position. Note that if we neglect the noise contribution given by large-scale structures, i.e. we assume the noise to be white, this lensing filter and the one for the galaxy overdensities are fully equivalent, except of course for the specific nature of the two.

As in the case of the optical filter, the application of the weak lensing filter to the data is done evaluating $A(\bmath \theta)$  over a grid of angular positions. Equation (\ref{lenssign}) is then discretized to be applied to real data. It becomes
\begin{equation}\label{lensdiscr}
A(\bmath \theta) = \frac{1}{n_g}\sum_{k=1}^{N} \epsilon_{tk} \Psi (|\bmath \theta_k - \bmath \theta|)\;,
\end{equation}
where $n_g$ is the number density of galaxies used for weak lensing measurement and $\epsilon_{tk}$ denotes the tangential component of the $k$-th galaxy ellipticity with respect to the angular position. Thus, the filtering is performed in the real domain, which requires to back-Fourier transform the filter function defined in Equation (\ref{lensfil}). For every estimate of $A$, its error $\sigma_A^2$ is also computed, according to Equation (\ref{sigma_A}). Peaks with S/N $>$ 3 are selected from the map and considered as detections.

\section{COSMOS field}

We now apply the optical filter and the weak lensing filter to the COSMOS field, a 2 square degree equatorial field that has been observed with the Hubble Space Telescope and with many other instruments, covering wavelengths from X-ray to radio \citep{cosmos}.

\subsection {Optical detections}\label{optcosm}

For the optical analysis, we use the public galaxy catalogue
with photometric redshifts \citep{b6}, considering the angular position, the photometric redshift, and the $i'$-band magnitude measured with Subaru \citep{b22}. The photometric redshifts presented in the catalogue were obtained from the galaxy fluxes in 30 different bands, ranging from the ultraviolet to the near infrared. To each galaxy a SED has been assigned during
the determination of the photometric redshift. We use this additional information
to calculate the k-correction. We search for clusters
between $z = 0.1$ and $z = 0.8$, with steps of $\Delta z =
0.02$. The upper limit of $z$ = 0.8 is motivated by two factors:
first, the model that we assume for the cluster luminosity function is based on low-redshift
objects, and cannot be extrapolated to too high redshifts; second, if the k-correction becomes
too strong for a given type of galaxies, our sample can no longer be assumed
complete up to a given rest-frame magnitude.
We scan an area smaller than the full field of view, such that, for
each redshift, the filter (cut at $r$ = 1 Mpc/$h$) is completely inside
the survey field. In this way we avoid any border effect, at the price
of loosing some possible detections of structures that have their
centre inside the COSMOS area, but which are not completely contained
in the field of view.

\begin{figure}
\begin{centering}
\includegraphics[width = 200 pt]{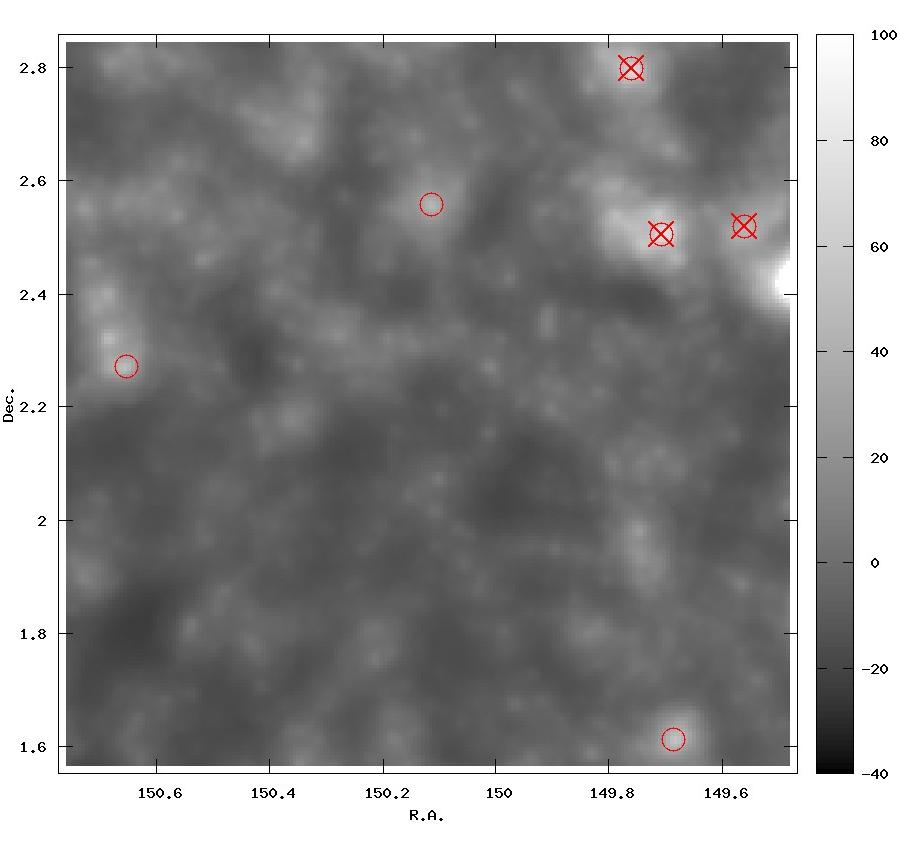}
\includegraphics[width = 200 pt]{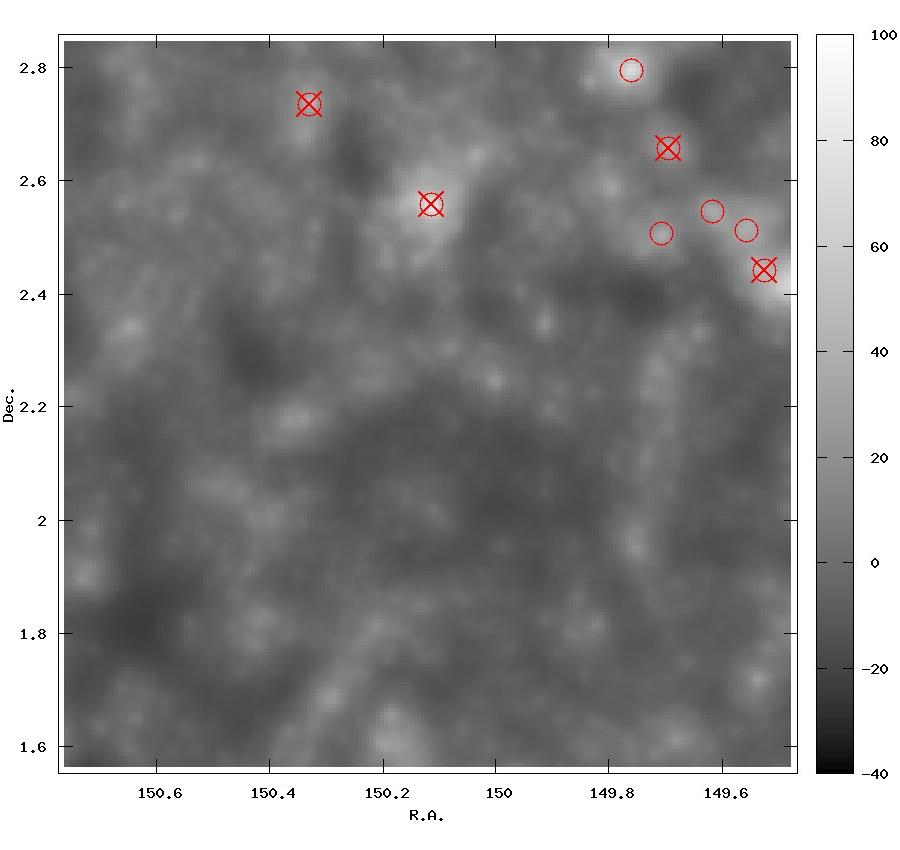}
\includegraphics[width = 200 pt]{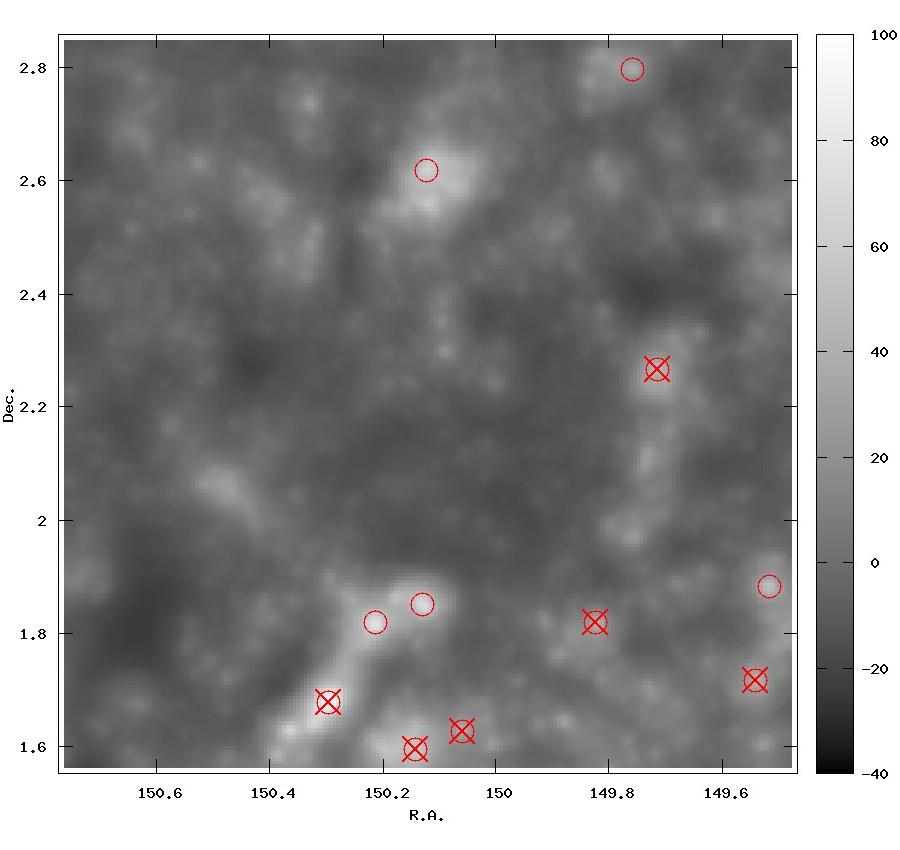}
\end{centering}
\caption{Maps of the estimates of the richness $\Lambda$ on the COSMOS field, at redshifts $z$
= 0.48, 0.5, 0.52, from top to bottom. Red circles indicate significant
peaks, while red crosses represent a cluster detection at \textit{that} redshift.}
\label{cosmos_05}
\end{figure}

In Fig.\ref{cosmos_05}, as examples, we show the results for the
slices at $z = 0.48$, $0.50$, $0.52$ where the detections with S/N $>$ 3 are marked
with circles.
Note that some structures are visible in more than one slice. In that case we assign as redshift of the detection the one where the likelihood is maximal, as discussed is Section \ref{reddet}.
In this way we obtain 140 significant detections with a redshift
estimate.

\begin{figure}
\begin{centering}
\includegraphics[scale = .65]{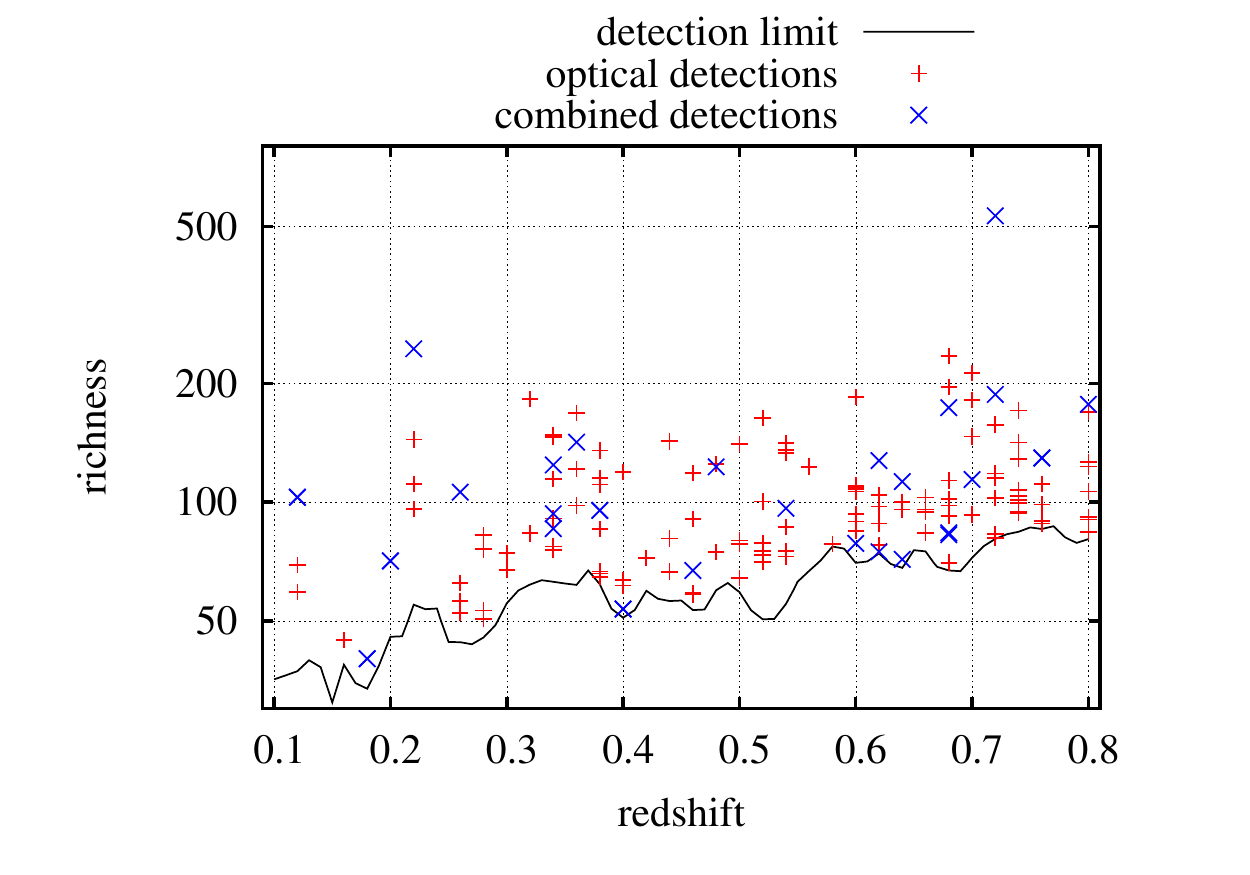}
\end{centering}
\caption{Redshift and (corrected) richness of the detected clusters.
The black line represents the detection limit, that depends on model and field
distributions at different redshifts. Clusters found in the optical analysis only are marked in red, those with a counterpart in the lensing analysis are marked in blue.}
\label{rich_red}
\end{figure}

Since the absolute magnitude limit changes with redshift, the richness
of clusters with the same galactic population is redshift dependent.
In particular the richness decreases as the redshift increases. To
correct for this effect and to obtain a quantity which depends on the
cluster galaxy content only, we make use of the luminosity function
$\phi(m)$ of a cluster at an arbitrary redshift $z_c = 0.2$ to compute the
normalization factor

\begin{equation}
R(z) = \int_0^{m_{lim}(z)} \phi(m)dm\;.
\end{equation}
This quantity represents the part of $\phi(m)$ visible at a given redshift $z$.
Dividing the measured richness $\Lambda$ for $R(z)$, where $z$ is the estimated
redshift of the cluster, we obtain a value that is directly
proportional to the physical galactic content of a cluster. More
precisely this \textquoteleft corrected richness' corresponds to the number of
cluster galaxies that would be visible if the cluster was located at
$z=0.2$. The \textquoteleft corrected richness' of our detections is shown in
Fig. \ref{rich_red}, together with the selection threshold S/N = 3
for optical detections already derived and shown in the lower panel of
Fig. \ref{sn}. Notice that the hills and wells of the curve are due to the change
of the field population as a function of redshift.

\subsection{Lensing detections}

The weak lensing cluster detections in COSMOS are performed by applying Equation (\ref{lensdiscr}) as discussed
by \citet{b7} on the shear catalogue obtained with the Suprime-Cam mounted on SUBARU
\citep{b21}. The galaxy number density in the catalogue is 42 $\rm arcmin^{-2}$. Following \citet{b21}, we assume that the mean redshift of the sources is 0.8. The weak lensing cluster model used in the filter construction is the signal expected from a cluster of mass $5 \times 10^{14} M_\odot$ at $z$ = 0.5.

The total number of significant (e.g. with S/N $>$ 3) detections is 82. The expected number of spurious detections at this S/N level is around 40\%, as computed by \cite{b9}. The derived weak lensing detections are related to the optical overdensities using a correlation length of 500 kpc/$h$. This quite
severe constraint is justified by the absence of any redshift
information in the lensing data. The 27 detections satisfying this correlation criterion are listed in
Table \ref{catalogue}, while optical detections without a lensing counterpart are listed in the Appendix.

In Fig. \ref{red_distr} we show the redshift distribution of our combined detections. From the source redshift distribution it is possible to evaluate the redshift sensitivity function for weak lensing, that is a measurement of how much the shear of observed galaxies is sensitive to structures at different redshifts.
In Fig. \ref{red_distr} we also show the optical sensitivity function, computed as the S/N of a cluster with 100 visible galaxies at $z = 0.2$. The ratio of lensing-confirmed clusters over the total number of optical detections in our analysis does not show any clear trend with redshift, but this could be due to the poor statistics.

\begin{figure}
\begin{centering}
\includegraphics[scale = 0.7]{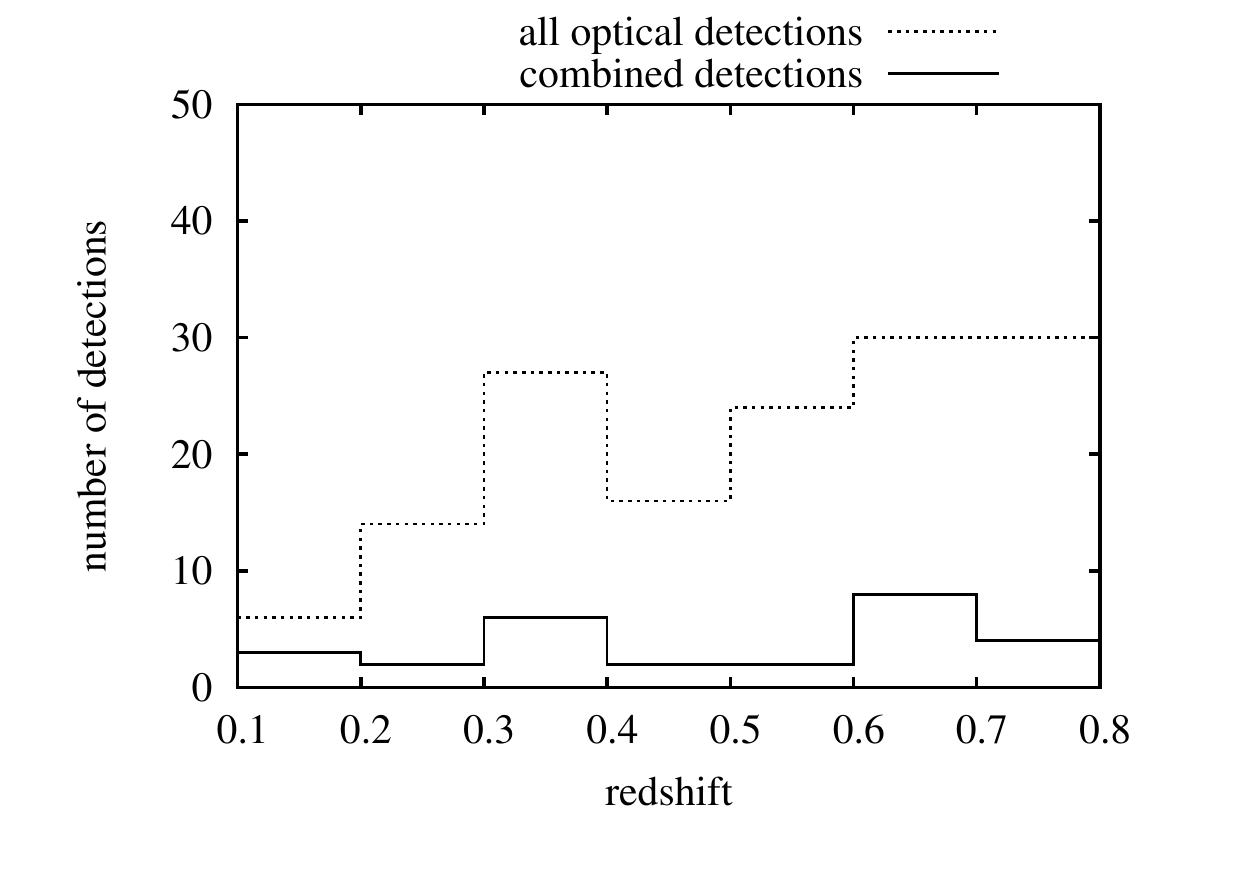}
\includegraphics[scale = 0.7]{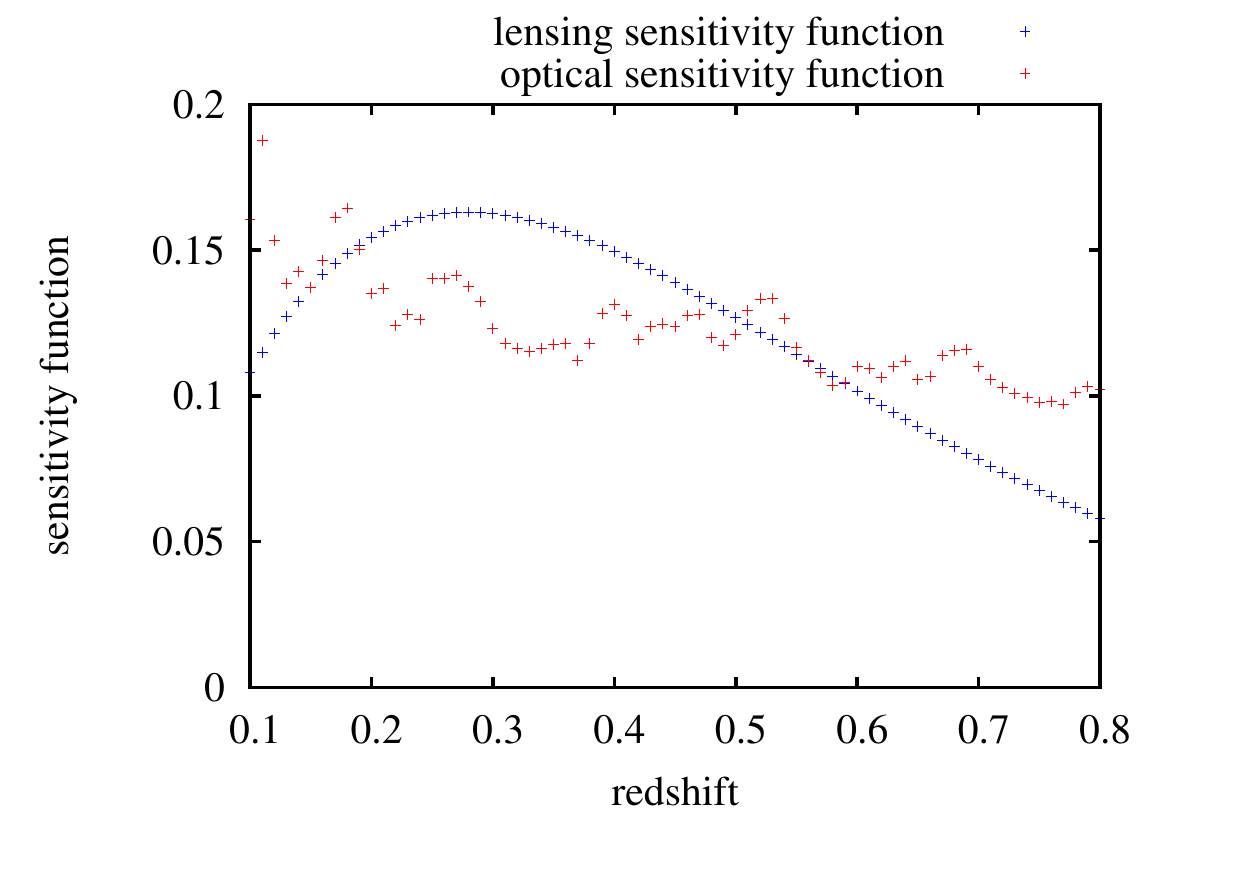}
\end{centering}
\caption{Top panel: redshift distribution of all optical detections (dashed line) and of the lensing-confirmed clusters (solid line). Bottom panel: lensing sensitivity
function and optical sensitivity function, defined as the S/N of a cluster
with 100 visible galaxies at $z = 0.2$, in arbitrary units. }
\label{red_distr}
\end{figure}

\subsection{Comparison with literature}

We can compare our results to published catalogues of clusters previously extracted from the COSMOS field. Previous analyses were done by \citet{b2}, combining optical and X-ray information and by
\citet{b8}, who found galaxy overdensities, by applying a Postman-like filter to
a previous version of the galaxy catalogue with no redshift information.
\citet{b15} analysed the COSMOS field with a wavelet method, and presented a
catalogue of large-scale structures. We note that these structures are in general much larger than our
targets, and that their size in some cases exceeds 10 Mpc in comoving coordinates. The few previous cluster detections in this
field via lensing were reported by \citet{b4}, while comparing the
performance of ground-based and space-based telescopes for weak
lensing, and by \citet{b21}.

Correlating our detections with results from literature, we use
different criteria depending on the characteristics of the
catalogues. For sources presented in \citet{b2} and \citet{b8}, we
match detections using a physical correlation length of 1 Mpc/$h$,
i.e. the cut-off applied to our radial filter. Instead, for the
larger structures presented in \citet{b15}, we use a correlation
length equal to the FWHM of their density peaks. In all these cases, the
correlation along the line-of-sight is performed considering
differences in redshift smaller than 0.05. This can be quite severe
for \citet{b8}, as their redshift determination was based on single
band observations without any redshift information. For the lensing detections presented by \citet{b4} and \citet{b21}, we use the same criterion we adopted for the internal matching with our lensing detections, i.e. a correlation length of 500 kpc/$h$. Regarding
\citet{b2}, we must stress that the comparison can only be done with
their catalogue of X-ray confirmed clusters, obtained correlating 420 optical overdensities with 150 X-ray diffuse sources. This justifies some discrepancies between the two analyses.

The existence of previous detections of the objects found in our optical and lensing analysis is reported in the last columns of Tables \ref{catalogue} and \ref{alldet}.

\begin{table}
\centering

\begin{tabular}{llllllc}
\hline
ID & $z$ & R.A. & Decl. &  S/N & richness & previous\\
&  & & &  & & detections \\
\hline
1 &  0.12  &    150.395  &      2.446  &      5.280  &    102.89  &  X \\
2 &  0.18  &    150.060  &      2.200  &      3.327  &     40.11  &   \\
3 &  0.20  &    150.616  &      2.425  &      3.802  &     70.95  &  X \\
4 &  0.22  &    150.192  &      1.650  &      6.628  &    244.74  &  XL \\
5 &  0.26  &    149.911  &      2.603  &      4.832  &    105.93  &  XSO \\
6 &  0.34  &    149.939  &      2.607  &      4.375  &    124.31  &  O \\
7 &  0.34  &    150.193  &      1.658  &      3.573  &     85.70  &  SL \\
8 &  0.34  &    150.299  &      1.609  &      3.749  &     93.49  &  XS \\
9 &  0.36  &    149.890  &      2.452  &      4.770  &    141.92  &  XS \\
10 &  0.38  &    150.142  &      2.053  &      3.831  &     95.29  &   \\
11 &  0.40  &    150.646  &      2.807  &      3.089  &     53.56  &   \\
12 &  0.46  &    150.687  &      2.400  &      3.419  &     67.07  &   \\
13 &  0.48  &    149.760  &      2.798  &      4.473  &    122.87  &  X \\
14 &  0.54  &    149.521  &      1.884  &      4.131  &     96.47  &   \\
15 &  0.60  &    149.926  &      2.519  &      3.202  &     78.57  &  L \\
16 &  0.62  &    149.572  &      1.882  &      3.018  &     74.84  &  S \\
17 &  0.62  &    150.590  &      2.473  &      4.035  &    127.47  &   \\
18 &  0.64  &    149.628  &      1.906  &      3.989  &    112.74  &  S \\
19 &  0.64  &    150.443  &      1.883  &      3.089  &     71.56  &   \\
20 &  0.68  &    149.718  &      1.816  &      3.428  &     82.65  &   \\
21 &  0.68  &    150.088  &      2.193  &      5.331  &    173.71  &  O \\
22 &  0.68  &    150.290  &      1.580  &      3.453  &     83.61  &   \\
23 &  0.70  &    150.308  &      2.406  &      3.975  &    114.17  &  S \\
24 &  0.72  &    149.921  &      2.521  &      8.721  &    532.42  &  XL \\
25 &  0.72  &    150.141  &      2.069  &      4.953  &    187.60  &   \\
26 &  0.76  &    150.641  &      2.804  &      3.764  &    129.46  &  S \\
27 &  0.80  &    150.437  &      2.763  &      4.705  &    177.08  &  S \\

\hline
\end{tabular}
\caption{Catalogue of the clusters detected with both optical and lensing
filters. The angular position is the one of the detection from optical data. In the last column we indicate whether the cluster has been found
previously with other techniques: \textquoteleft X' stands for \citet{b2}, \textquoteleft S' for \citet{b15}, \textquoteleft O' for \citet{b8}, \textquoteleft L' for \citet{b4} or \citet{b21}.}
\label{catalogue}
\end{table}

We note here the ability of our optical filter to distinguish different galaxy clusters that are aligned along the line of sight, despite their signals are degenerate in weak lensing analyses. In fact, the two detections with identification numbers 4 and 7 correspond to SLJ1000.7+0137, a weak lensing and X-ray source studied by \citet{ham} with spectroscopical observations. They found that it is actually made by two, or possibly three, overdensities of galaxies at different redshifts, almost at the same angular position. Our method is able to disentagle the two structures at $z = 0.22$ and $z = 0.34$ using information from photometric redshifts only. A similar case is that of SLJ1001.2+0135, described by \citet{ham} as the superposition of two overdensities, located at $z = 0.22$ and $z = 0.37$. Our detection 8 corresponds to the latter, while at lower redshifts that angular position is outside the scanned area, because the filter would not be completely contained in the field of view (see Section \ref{optcosm}). However, if we force our algorithm to analyse that region, we find an overdensity at $z = 0.24$ with coordinates (R.A. = 150.329, Dec. = 1.609), confirming the alignment of two different structures.

\subsection{Characterizing our sample}

\begin{figure}
\begin{centering}
\includegraphics[scale = 0.7]{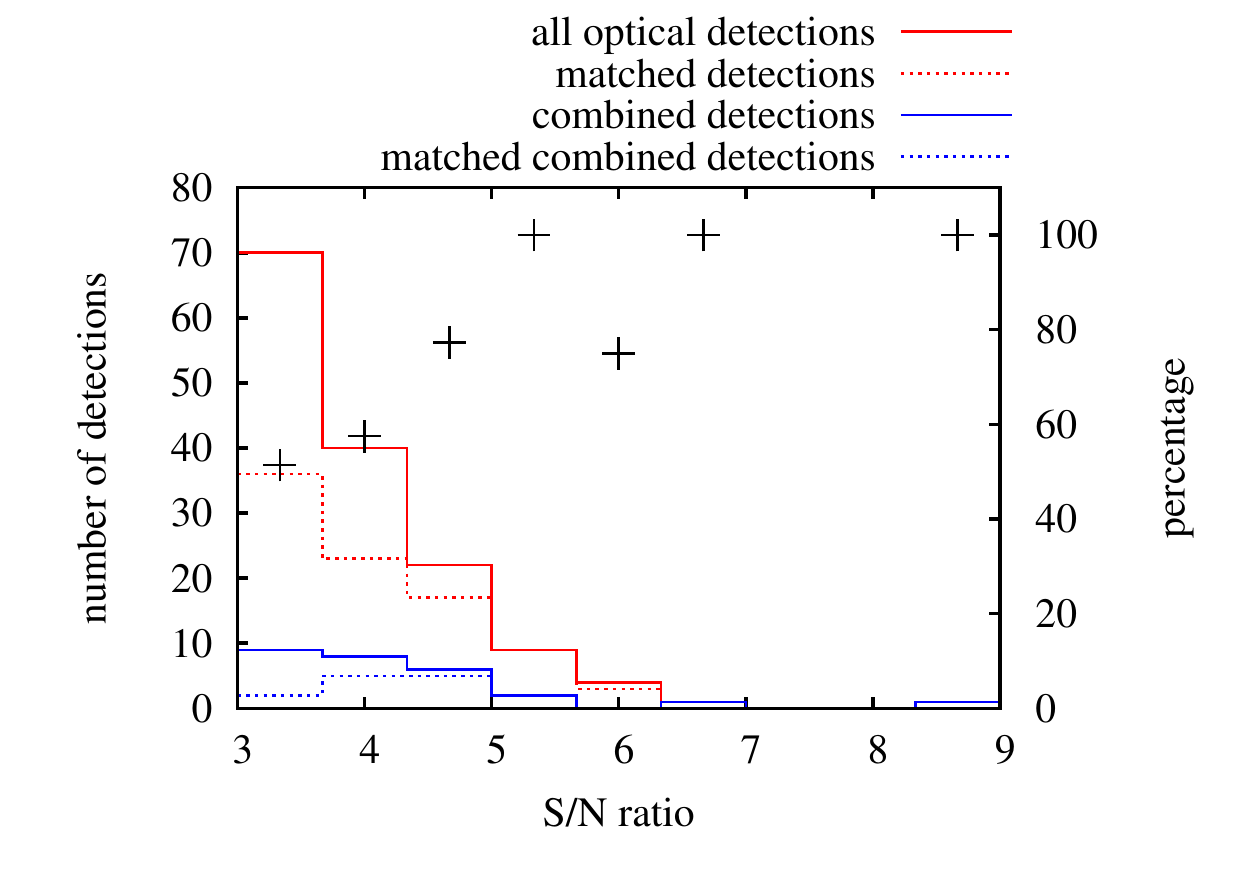}
\end{centering}
\caption{S/N distribution of all the significant optical detections (solid red), and of the detections with a counterpart in other analyses (dotted red). In blue we show the same distributions, but considering only combined (optical+lensing) detections. The black crosses indicate the percentage of our optical detections having at least one counterpart (scale on the right).}
\label{sn_distr}
\end{figure}

In Fig.\ref{sn_distr} we plot the number of detections obtained with
our analysis (both optical and optical plus lensing, dashed and solid lines respectively), as a function of their optical S/N ratio together with the
corresponding amount of detections already reported in literature.
As expected, the percentage of \textquoteleft matched' detections
increases as a function of S/N. The only detection with S/N $>$ 5
without a counterpart is actually near to a large structure observed
by \citet{b15} with a redshift mismatch of $\Delta z = 0.06$, only
slightly larger than the limit we adopted.

\begin{figure}%[t!hb]
\begin{centering}
\includegraphics[scale = 0.7]{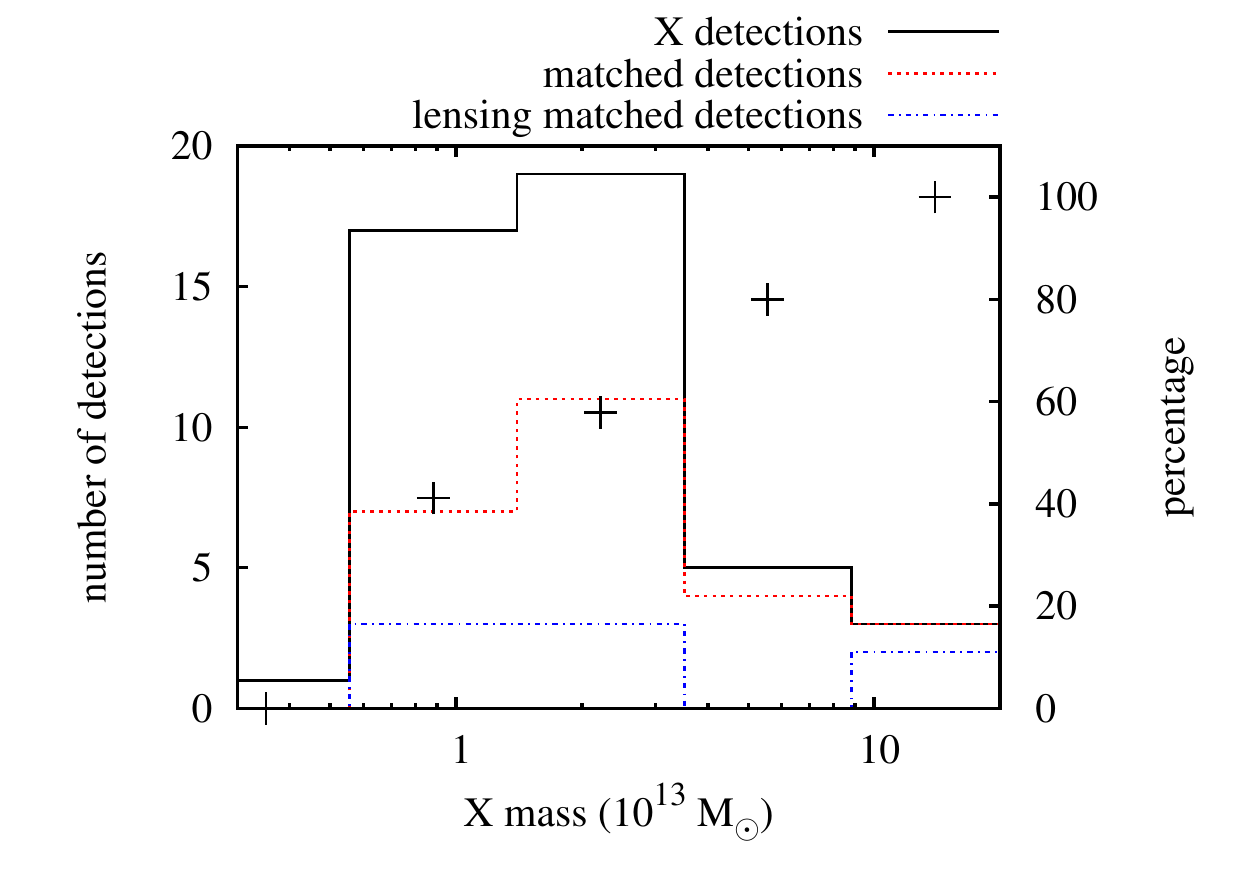}
\end{centering}
\caption{Mass distribution of all the X-ray detected clusters (in black), of the clusters we detect in the optical analysis (in red), and of those we detect in both optical and lensing analyses (in blue). The black crosses indicates the percentage of clusters detected in our optical analysis.}
\label{x_distr}
\end{figure}

To investigate the selection function of our method with respect to
the mass, we plot in Fig. \ref{x_distr} the number of
detections with respect to the X-ray masses estimated by \citet{b2}. We
include in this analysis only those X-ray clusters which lie inside the
volume of our search, avoiding borders of the field and too high
redshifts, as already
discussed in Section \ref{optcosm}. We see that the rate of X-ray clusters that
are optically confirmed by our analysis increases with their mass, and it is above 50\% for clusters of mass above $1.5 \times 10^{13} M_\odot$.

\begin{figure}%[htb]
\begin{centering}
\includegraphics[scale = 0.7]{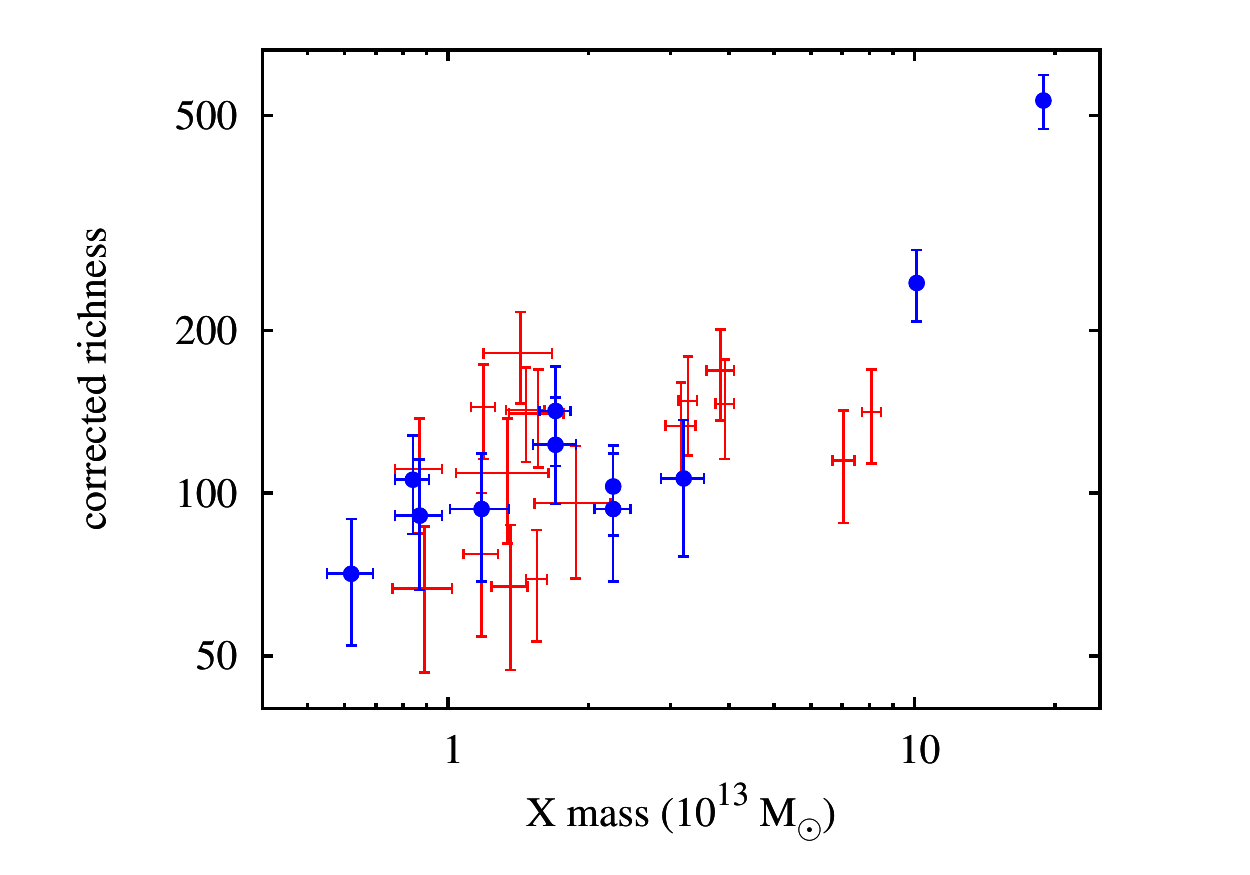}
\end{centering}
\caption{Correlation between the X-ray mass derived by
\citet{b2} and our corrected richness. In red we show all the optical detections, in blue the lensing-confirmed clusters.}
\label{X_lensrich}
\end{figure}

Finally we compare in Fig. \ref{X_lensrich} the corrected richness, corresponding to a
measurement of the mass in galaxies, with the X-ray mass estimates
made by \citet{b2}. The proportionality
between the two \textquoteleft mass estimates' is clear although the scatter is very
large. Note that the corrected richness is only a relative measure of the cluster stellar mass, in the sense that we do not quantify the physical mass
scale. Note also that
our filter has a fixed physical size, while the X-ray analysis was
done inside an estimated $r_{500}$ for each cluster, partially explaining the differences. The noise in the richness estimate for low-mass clusters is likely to be due to the field galaxies we observe inside our fixed spatial filter. This is expected given the error bars that show our analytic estimate.

\section{Summary and conclusions}

In this paper we have presented an optimal linear filtering
technique for the detection of galaxy clusters from optical and weak
lensing data. The filter was first presented by \citet{b7} for weak
lensing analysis and we now use the same formalism to extend it to
optical data. The filter relies on a physical model for clusters
and thus it accounts for all the known properties of galaxy clusters
which can be inferred from photometric observations.
Different information can be included and in this work we decided to
restrict our analysis to the following observed properties of the
galaxies: positions, magnitudes in one band, and photometric redshifts if present.
The algorithm starts from a galaxy catalogue with the mentioned physical quantities and creates a map of values for the richness parameter $\Lambda$ at different redshifts. The peaks of these maps represent the possible locations of galaxy clusters. The redshift of the clusters is determined by
maximizing the likelihood of the data as a function of the redshift of the cluster model.

Our algorithm offers a number of improvements over existing filtering methods. Photometric redshifts are included in a flexible way, that adapts itself to the precision of the measurement of each galaxy, avoiding any sharp cut of the catalogue in redshift slices. It uses results from observed galaxy clusters to build a suitable model for cluster galaxy distributions. The richness calculated by the algorithm at the position of the cluster is
a measurement of the number of galaxies in the detected cluster, that can be corrected for redshift dimming. Moreover, from the expected spatial profile and luminosity distribution of the clusters, and from the observed field population, it is
possible to evaluate the noise of the detection and its significance. We tested the algorithm with numerical simulations, probing that it is able to obtain an unbiased estimate of the richness of the cluster, with an uncertainty
predicted analytically from the model, and also its redshift.

We applied both the weak lensing and the optical filters to real data,
obtaining a catalogue of candidate galaxy clusters for the COSMOS field. We presented a catalogue of 27 lensing-confirmed clusters, 11 of which do not have any previous detection in the literature. For the subsample of objects that have been analysed through their X-ray emission, we found that we are able to detect more than $50\%$ of the clusters with an X-ray mass over $1.5 \times 10^{13}$ $M_\odot$ and we found a good correlation between our galaxy richness parameter and the cluster mass determined from X-ray temperature.

\appendix

\section{Optical-only detections}
In Table \ref{alldet} we list the optical detections without lensing
counterpart.

\begin{table}
\centering
\begin{tabular}{llllllc}
\hline
ID & $z$ & R.A. & Decl. &  S/N & richness & previous\\
&  & & &  &  & detections \\
\hline
 28  &    0.12  &     149.862  &       1.765  &       4.279  &
 69.35  &  X \\
 29  &    0.12  &     150.432  &       2.630  &       3.925  &
 59.13  &   \\
 30  &    0.16  &     150.387  &       2.069  &       3.246  &
 44.64  &   \\
 31  &    0.22  &     150.102  &       2.358  &       5.042  &
 144.29  &  XSO \\
 32  &    0.22  &     150.395  &       2.460  &       4.397  &
 111.20  &  O \\
 33  &    0.22  &     150.620  &       2.707  &       4.070  &
 96.12  &  S \\
 34  &    0.26  &     149.842  &       2.682  &       3.436  &
 56.19  &  S \\
 35  &    0.26  &     150.089  &       2.474  &       3.640  &
 62.42  &   \\
 36  &    0.26  &     150.268  &       2.672  &       3.305  &
 52.37  &  S \\
 37  &    0.28  &     149.594  &       2.150  &       3.978  &
 75.93  &   \\
 38  &    0.28  &     149.969  &       2.450  &       3.190  &
 50.52  &  S \\
 39  &    0.28  &     150.044  &       2.225  &       4.159  &
 82.51  &  SO \\
 40  &    0.28  &     150.194  &       1.747  &       3.280  &
 53.16  &  S \\
 41  &    0.30  &     150.186  &       2.794  &       3.517  &
 74.18  &   \\
 42  &    0.30  &     150.572  &       1.942  &       3.337  &
 67.24  &  X \\
 43  &    0.32  &     150.456  &       2.049  &       5.332  &
 182.39  &  S \\
 44  &    0.32  &     150.652  &       2.313  &       3.531  &
 83.32  &  S \\
 45  &    0.34  &     149.596  &       2.820  &       4.773  &
 146.29  &  XS \\
 46  &    0.34  &     149.767  &       2.329  &       3.336  &
 75.73  &  S \\
 47  &    0.34  &     150.070  &       2.378  &       3.372  &
 77.19  &  X \\
 48  &    0.34  &     150.185  &       1.765  &       4.804  &
 148.11  &  XS \\
 49  &    0.34  &     150.373  &       2.444  &       3.691  &
 90.87  &  XS \\
 50  &    0.34  &     150.536  &       2.730  &       4.186  &
 114.54  &  S \\
 51  &    0.36  &     149.787  &       2.167  &       3.900  &
 98.05  &  S \\
 52  &    0.36  &     150.119  &       2.689  &       5.227  &
 168.45  &  X \\
 53  &    0.36  &     150.523  &       2.570  &       4.385  &
 121.46  &   \\
 54  &    0.38  &     149.675  &       2.413  &       3.133  &
 66.70  &  S \\
 55  &    0.38  &     149.767  &       1.625  &       3.112  &
 65.94  &   \\
 56  &    0.38  &     149.821  &       2.275  &       4.637  &
 135.17  &  SO \\
 57  &    0.38  &     149.966  &       1.678  &       4.250  &
 115.08  &  X \\
 58  &    0.38  &     150.234  &       2.474  &       3.609  &
 85.62  &   \\
 59  &    0.38  &     150.387  &       2.413  &       4.162  &
 110.80  &  X \\
 60  &    0.38  &     150.663  &       2.559  &       3.077  &
 64.64  &   \\
 61  &    0.40  &     149.578  &       2.533  &       3.343  &
 61.52  &   \\
 62  &    0.40  &     149.971  &       2.748  &       3.401  &
 63.40  &   \\
 63  &    0.40  &     150.356  &       2.651  &       4.816  &
 119.47  &   \\
 64  &    0.42  &     149.674  &       2.731  &       3.336  &
 72.20  &   \\
 65  &    0.44  &     149.961  &       2.210  &       3.315  &
 66.64  &  XO \\
 66  &    0.44  &     150.494  &       2.070  &       5.034  &
 142.59  &  XSO \\
 67  &    0.44  &     150.690  &       2.021  &       3.695  &
 80.85  &   \\
 68  &    0.46  &     149.667  &       1.625  &       4.658  &
 118.42  &  S \\
 69  &    0.46  &     150.354  &       2.741  &       3.178  &
 58.86  &   \\
 70  &    0.46  &     150.653  &       2.271  &       4.031  &
 90.50  &   \\
 71  &    0.46  &     150.721  &       2.244  &       3.160  &
 58.26  &   \\
 72  &    0.48  &     149.561  &       2.519  &       3.407  &
 74.82  &   \\
 73  &    0.48  &     149.707  &       2.506  &       4.519  &
 125.21  &  S \\
 74  &    0.50  &     149.525  &       2.441  &      3.519  &
 78.18  &  S \\
 75  &    0.50  &     149.695  &       2.657  &       3.148  &
 64.27  &   \\
 76  &    0.50  &     150.113  &       2.559  &       4.855  &
 140.51  &  XS \\
 77  &    0.50  &     150.328  &       2.735  &       3.562  &
 79.87  &  S \\
 78  &    0.52  &     149.542  &       1.719  &       3.733  &
 73.37  &   \\
 79  &    0.52  &     149.714  &       2.267  &       3.885  &
 78.72  &  O \\
 80  &    0.52  &     149.823  &       1.821  &       3.785  &
 75.19  &   \\
 81  &    0.52  &     150.059  &       1.629  &       3.646  &
 70.42  &   \\
 82  &    0.52  &     150.141  &       1.597  &       4.453  &
 100.43  &   \\
 83  &    0.52  &     150.295  &       1.680  &       5.809  &
 163.77  &   \\
 84  &    0.54  &     150.135  &       1.853  &       4.970  &
 135.34  &   \\
 85  &    0.54  &     150.216  &       1.821  &       4.928  &
 133.21  &  X \\
 86  &    0.54  &     150.260  &       1.765  &       3.528  &
 72.84  &   \\
 87  &    0.54  &     150.360  &       1.627  &       3.883  &
 86.33  &   \\
 88  &    0.54  &     150.467  &       2.066  &       5.088  &
 141.32  &  X \\
 89  &    0.54  &     150.573  &       2.166  &       3.591  &
 75.14  &   \\
 \hline
\end{tabular}
\end{table}
\begin{table}
\centering
\begin{tabular}{llllllc}
\hline
ID & $z$ & R.A. & Decl. &  S/N & richness & previous\\
&  & & &  &  & detections \\
\hline
 90  &    0.56  &     149.525  &       1.760  &       4.173  &
 123.00  &   \\
 91  &    0.58  &     149.578  &       1.683  &       3.028  &
 78.39  &  S \\
 92  &    0.60  &     149.617  &       1.740  &       5.134  &
 184.96  &  S \\
 93  &    0.60  &     150.063  &       2.798  &       3.874  &
 110.18  &   \\
 94  &    0.60  &     150.253  &       2.346  &       3.334  &
 84.34  &   \\
 95  &    0.60  &     150.301  &       2.804  &       3.445  &
 89.34  &   \\
 96  &    0.60  &     150.491  &       2.745  &       3.851  &
 109.03  &  X \\
 97  &    0.60  &     150.574  &       2.608  &       3.827  &
 107.78  &   \\
 98  &    0.60  &     150.616  &       2.780  &       3.527  &
 93.14  &   \\
 99  &    0.60  &     150.717  &       2.537  &       3.799  &
 106.39  &   \\
100  &    0.62  &     150.052  &       2.321  &       3.082  &
77.73  &   \\
101  &    0.62  &     150.520  &       2.473  &       3.620  &
104.23  &   \\
102  &    0.62  &     150.707  &       2.759  &       3.306  &
88.25  &   \\
103  &    0.62  &     150.725  &       2.625  &       3.492  &
97.54  &   \\
104  &    0.64  &     150.633  &       2.715  &       3.738  &
100.25  &   \\
105  &    0.64  &     150.737  &       2.825  &       3.644  &
95.81  &  X \\
106  &    0.66  &     149.802  &       1.804  &       3.204  &
83.66  &  S \\
107  &    0.66  &     150.185  &       2.158  &       3.438  &
94.33  &   \\
108  &    0.66  &     150.196  &       2.238  &       3.613  &
102.74  &   \\
109  &    0.66  &     150.687  &       1.667  &       3.469  &
95.77  &   \\
110  &    0.68  &     149.673  &       2.277  &       3.670  &
92.24  &   \\
111  &    0.68  &     149.925  &       2.597  &       3.092  &
70.21  &  S \\
112  &    0.68  &     149.948  &       2.097  &       4.162  &
113.53  &   \\
113  &    0.68  &     150.010  &       2.119  &       3.810  &
98.07  &   \\
114  &    0.68  &     150.060  &       2.608  &       6.305  &
234.42  &  SO \\
115  &    0.68  &     150.173  &       2.518  &       5.706  &
195.95  &  S \\
116  &    0.68  &     150.257  &       1.968  &       3.895  &
101.70  &   \\
117  &    0.70  &     149.647  &       2.828  &       3.514  &
93.08  &   \\
118  &    0.70  &     149.964  &       2.673  &       4.606  &
146.79  &  S \\
119  &    0.70  &     149.986  &       2.578  &       5.200  &
181.54  &  XSO \\
120  &    0.70  &     150.003  &       2.451  &       5.684  &
212.81  &  S \\
121  &    0.70  &     150.152  &       2.601  &       3.509  &
 92.83  &  S \\
122  &    0.72  &     149.899  &       2.394  &       3.789  &
118.18  &  S \\
123  &    0.72  &     150.086  &       2.460  &       4.479  &
157.29  &  S \\
124  &    0.72  &     150.108  &       2.565  &       3.475  &
102.36  &  S \\
125  &    0.72  &     150.207  &       2.361  &       3.056  &
83.12  &  S \\
126  &    0.72  &     150.593  &       2.129  &       3.011  &
81.17  &   \\
127  &    0.72  &     150.736  &       2.416  &       3.731  &
115.19  &  S \\
128  &    0.74  &     149.523  &       2.656  &       3.344  &
101.46  &   \\
129  &    0.74  &     149.866  &       2.492  &       4.466  &
170.82  &  S \\
130  &    0.74  &     149.986  &       2.563  &       3.305  &
99.40  &  SO \\
131  &    0.74  &     150.024  &       2.688  &       3.388  &
103.83  &  S \\
132  &    0.74  &     150.051  &       2.302  &       3.821  &
128.61  &  S \\
133  &    0.74  &     150.116  &       2.705  &       4.030  &
141.57  &  SO \\
134  &    0.74  &     150.171  &       1.703  &       3.449  &
107.13  &   \\
135  &    0.74  &     150.296  &       2.378  &       3.196  &
93.72  &  S \\
136  &    0.74  &     150.547  &       2.803  &       3.212  &
94.55  &  S \\
137  &    0.76  &     149.559  &       1.647  &       3.054  &
88.26  &   \\
138  &    0.76  &     149.732  &       2.761  &       3.248  &
98.66  &   \\
139  &    0.76  &     149.839  &       1.684  &       3.468  &
111.29  &   \\
140  &    0.76  &     150.114  &       2.255  &       3.080  &
89.62  &  S \\
141  &    0.80  &     149.706  &       2.265  &       3.251  &
91.47  &   \\
142  &    0.80  &     150.040  &       2.652  &       3.906  &
126.31  &   \\
143  &    0.80  &     150.368  &       2.005  &       3.546  &
106.42  &  X \\
144  &    0.80  &     150.506  &       2.222  &       3.229  &
90.37  &   \\
145  &    0.80  &     150.538  &       2.148  &       3.097  &
84.12  &  O \\
146  &    0.80  &     150.580  &       2.652  &       4.594  &
169.49  &  S \\
147  &    0.80  &     150.702  &       2.769  &       3.850  &
123.09  &   \\
\hline
\end{tabular}
\caption{As Table 1, but for clusters detected with the optical filter without lensing counterpart.}
\label{alldet}
\end{table}

\section*{Acknowledgments}

We acknowledge financial contributions from contracts ASI-INAF I/023/05/0, ASI-INAF I/088/06/0, and ASI \textquoteleft EUCLID-DUNE' I/064/08/0. This work was supported by the Transregio-Sonderforschungsbereich TR 33 of the Deutsche Forschungsgemeinschaft. FB and TH thank the Institut f\"{u}r Theoretische Astrophysik of the University of Heidelberg for hospitality during the preparation of this work. We are grateful to M. Bartelmann for having carefully read the paper and provided useful comments. We also thank the anonymous referee for her/his interesting and stimulating remarks.

\label{lastpage}

\end{document}